\documentclass[journal]{IEEEtran}
\pdfoutput = 1
\usepackage[font=footnotesize]{caption}
\usepackage{graphicx}
\usepackage{subcaption}
\usepackage{amsthm}
\usepackage{mathrsfs}  
\usepackage{amsmath,amssymb,amsthm,amsbsy}
\usepackage{mathtools}
\usepackage{commath}
\usepackage{breqn}
\usepackage{xcolor}
\usepackage{url}
\usepackage{float}
\usepackage[style = ieee]{biblatex}
\usepackage{units}
 
\usepackage{cuted}
\usepackage{makecell}
\usepackage{multirow}
\usepackage{float}
\usepackage{footnote}
\makesavenoteenv{tabular}
\usepackage[utf8]{inputenc}
\usepackage[T1]{fontenc}
\usepackage{balance}
\usepackage{algorithm,algorithmic}

\newcommand{\kc}[1]{{\color{black}#1}}
\newcommand{\kcc}[1]{{\color{black}#1}}
\usepackage{steinmetz}
\usepackage{gensymb}

\usepackage[style = ieee]{biblatex}
\bibliography{main}

\ifCLASSINFOpdf
 
\else

\fi

\begin{document}

\title{\vspace{-0.3cm} Dissipation of Oscillation Energy and Distribution of Damping Power in a Multimachine Power System: A Small-signal Analysis}

\author{Kaustav~Chatterjee,~\IEEEmembership{Student Member,~IEEE}~~and
        ~Nilanjan~Ray~Chaudhuri,~\IEEEmembership{Senior~Member,~IEEE} \vspace{-0.5cm}
\thanks{K. Chatterjee and N. R. Chaudhuri are with The School
of Electrical Engineering and Computer Science, The Pennsylvania State University, University Park,
PA 16802, USA (e-mail: kuc760@psu.edu; nuc88@psu.edu)}

\thanks{Financial support from the NSF grant under award CNS 1739206 is gratefully acknowledged.}}

\maketitle

\begin{abstract}
This paper revisits the concept of damping torque in a multimachine power system and its relation to the dissipation of oscillation energy in synchronous machine windings. As a multimachine extension of an existing result on a single-machine-infinite-bus (SMIB) system, we show that  the total damping power for a mode stemming from the interaction of electromagnetic torques and rotor speeds is equal to the sum of average
power dissipations in the generator windings corresponding to the modal oscillation. Further, counter-intuitive to the SMIB result, we demonstrate that, although the equality holds on an aggregate,
such is not the case for individual machines in an interconnected system. To that end, distribution factors
 are derived for expressing the average damping power of each generator as a linear combination of average powers of modal energy dissipation in the windings of all machines in the system. These factors represent 
the distribution of damping power in a multimachine system. The results 
are validated on IEEE 4-machine and 16-machine test systems. 
\end{abstract}



\IEEEpeerreviewmaketitle

\section{Introduction}
\label{intro}
\IEEEPARstart{I}{n} small-signal analysis, stability of a power system under electromechanical oscillations is determined by studying the eigenvalues of the system linearized around the quasi-steady-state operating point or by performing modal estimation on the response variables. In either case, the damping ratios obtained indicate 
the margin of system stability, but do not quantify the damping contribution from individual sources. In that regard, the small-signal representation of a generator's electromagnetic torque as a phasor in its synchronously rotating rotor speed-angle reference frame offers  geometric intuition into decomposing the torque into its damping and synchronizing components. 
Conceptualized by Park in his 1933 paper \cite{Park} and furthered by Concordia \cite{Concordia2, concordia}, Shepherd \cite{shepherd}, and notable others \cite{kilgore, demello, alden1, alden2, shaltout3}, the damping and synchronizing torque coefficients contribute towards insightful understanding of the stabilizing contributions coming from the machine and its associated governor and excitation systems. Consequently, some of the early designs of power system stabilizers for damping oscillations have evolved out of these notions. However, historically, these studies on damping torque have been presented either considering a SMIB system or by reducing the system to a linearized single-machine equivalent. 
Unlike a SMIB system, damping torque of a generator in a multimachine system depends not only on its own speed-deviation but also on that of the other machines, their excitation systems, and the overall network structure and parameters. 
Also, the analytical modeling of these differ in literature, for instance, authors in \cite{chiang} model damping torque as a higher degree polynomial in speed-deviations, in contrast to linear terms in \cite{pai}.
The $1999$ IEEE task force report investigating modeling adequacy for representing damping in multimachine stability studies \cite{taskf} identified eight different sources of damping and 
recommended abstracting their contributions into a single retarding torque in the swing equation of each generator. 

Damping torque, thus, over the years, has largely remained a \textit{conceptual} tool  for analyzing stability in power systems.  
Although some of the initial works listed before highlighted an \textit{intuitive} link between damping torque and dissipation of oscillation energy, it is only in the recent works \cite{chen2} and \cite{tsingua1} that a rigorous mathematical connection between the two has been established for a SMIB system. However, for multimachine systems such a connection is yet to be confirmed 
-- in this paper, we 
make a maiden attempt to fill this gap. To that end, we use a simplified mathematical model for multimachine systems to establish an equivalence between the average power dissipation due to the 
damping torques on the rotors and the 
average rates of oscillation energy dissipation in the machine windings.
In this context, we make a note of \cite{cai}, where claims regarding the consistency of damping and dissipation coefficients in multimachine systems are made based on presupposition of this equality without any formal proof. 
Going ahead, in the paper we also demonstrate that 
the damping power of each machine stemming from the interaction of its own speed and torque, is derived in parts from the rates of energy dissipation in the windings of all machines, over and above it's own winding. 
 

\textcolor{black}{At this point, it is important to clarify that the focus of this paper is not on improving the algorithmic tools of [14] or [13] for perfecting the science of locating oscillation sources or to present an alternate path for doing the same, but to derive further analytical insights from the findings of these two seminal papers. Our primary intention is to bridge a mathematical connection between the concept of transient energy dissipation, as introduced in [14], and the notion of damping torque, which is nearly a century old.}

The contributions of the paper are as follows: (1) we develop a phasor-based small-signal-formulation for calculating the mode-wise average powers (i.e. damping powers)  of 
electromechanical oscillation due to the interaction of damping torque and speed in each machine; (2) using this framework for a simplified system model with lossless transmission network and constant power loads, we extend the SMIB results in \cite{tsingua1} to show that the total damping power for a mode 
is equal to the sum of average
power dissipations in the generator windings corresponding to the modal oscillation;
and finally, (3) 
counter-intuitive to the SMIB result in \cite{tsingua1}, we demonstrate that the aforementioned equality does not hold for individual machines in an interconnected system -- in fact the damping power in each machine can be expressed as a weighted linear combination of power dissipation in windings of different generators. 
These weighing factors (called `distribution factors' in the paper) are analytically derived $-$ which essentially describe the participation of the power dissipation in different machines in constituting the damping power of each generator.
 
To that end, in the next section, we derive a linearized representation of the simplified system mentioned earlier with a third-order synchronous generator model. Building on this model, we present contributions (1)$-$(3) in Sections \ref{secDP}-\ref{distri}, which are followed by case studies on IEEE 4-machine and 16-machine test systems in Section \ref{result} to validate the claims -- both for the simplified model used in derivation, and for systems with detailed machine models. Finally, concluding remarks are presented in Section VII.

\par \textit{Notations:} 
Superscripts $T$, $*$, and $H$ are respectively the transpose, conjugate, and Hermitian operators. $\Re\{\cdot\}$ and $\Im\{\cdot\}$ denote the real and imaginary parts of a complex entity.
\vspace{-5pt}
\section{Simplified System Model: Linearized Representation}
\label{model}
Consider a $n-$bus transmission system of which, without loss of generality, first $n_g$ are designated as generator buses. The network is lossless and each synchronous generator is described by a third-order machine model capturing the electromechanical dynamics of the rotor and the field flux. In addition, assume manual excitation for the generators and constant power loads at all buses. 
\par The differential equations describing the dynamics of each generator are same as those given in eqns (6.132) $-$ (6.134) of \cite{pai}. Further, the stator and network algebraic equations can be obtained from eqns (6.142) $-$ (6.144). Since we assume a lossless network, in eqns (6.143) and (6.144) of \cite{pai}, $\propto_{ik} = \pi/2$ for $i\neq k$, and $\propto_{ii} = -\pi/2$. Unless specified otherwise, all symbols have their usual meanings as in \cite{pai}. 

\par We eliminate the stator algebraic equations by substituting $I_{d_i}$ and $I_{q_i}$ obtained from eqn (6.142) into eqns (6.143) and (6.144) of \cite{pai} $-$ with stator resistances neglected. This leaves us with $3n_g$ differential equations and $2n$ algebraic equations as functions of state variables $\delta_i$, $\omega_i$, and $E_{q_i}'$, for $i = 1 \dots n_g$, and algebraic variables $\theta_i$ and $V_i$, for $ i = 1 \dots n$ as described below
\begin{equation}
\small
\label{f1} \dot \delta_i = \omega_i - \omega_s
\end{equation}
\begin{equation}
\small
\label{f2} \frac{\dot \omega_i}{\omega_s} = 
        \frac{T_{m_i}}{2H_i} - \frac{E_{q_i}'V_i\sin(\delta_i - \theta_i)}{2H_i~x_{d_i}'} + \frac{V_i^2\sin2(\delta_i-\theta_i)}{4H_i}\Big(\frac{x_{q_i} - x_{d_i}'}{x_{q_i}x_{d_i}'}\Big)
\end{equation}
\begin{equation}
\small
\begin{aligned}
\label{f3} \dot E_{q_i}' = \frac{E_{fd_i}}{T_{do_i}'} - \frac{E_{q_i}'}{T_{do_i}'} - &\frac{E_{q_i}' - V_i \cos(\delta_i-\theta_i)}{x_{d_i}'} \Big( \frac{x_{d_i}-x_{d_i}'}{T_{do_i}'} \Big) \\ 
\\ & \hspace{3cm}\text{for} ~i = 1,\dots n_g
\end{aligned}
\end{equation}
\begin{equation}
\small
\label{g1}0 = f_{i} = \begin{cases} \begin{split} 
    \frac{E_{q_i}'V_i\sin(\delta_i - \theta_i)}{x_{d_i}'} - \frac{V_i^2\sin2(\delta_i-\theta_i)}{2}\Big(\frac{x_{q_i} - x_{d_i}'}{x_{q_i}x_{d_i}'}\Big)\\ \hspace{-0.3cm} +~ P_{L_i} -\sum_{k=1, k\neq i}^n V_iV_kY_{ik}\sin(\theta_i-\theta_k)\\ \hspace{-0.3cm} \text{for} ~i = 1,\dots n_g \end{split} \\
    \begin{split}
       P_{L_i} -\sum_{k=1, k\neq i}^n V_iV_kY_{ik}\sin(\theta_i-\theta_k) \\  \text{for} ~i = n_g+1,\dots n
    \end{split}
\end{cases}
\end{equation}
\begin{equation}
\small
\label{g2} 0 = g_{i} = \begin{cases} \begin{split}
    \frac{E_{q_i}'V_i\cos(\delta_i - \theta_i)}{x_{d_i}'} - \frac{V_i^2\cos^2(\delta_i-\theta_i)}{x_{d_i}'} + Q_{L_i}- V_i^2Y_{ii} \\- \frac{V_i^2\sin^2(\delta_i-\theta_i)}{x_{q_i}} + \sum_{k=1, k\neq i}^n V_iV_kY_{ik}\cos(\theta_i-\theta_k)\\ \text{for} ~i = 1,\dots n_g \end{split} \\
    \begin{split}
       Q_{L_i} + \sum_{k=1, k\neq i}^n V_iV_kY_{ik}\cos(\theta_i-\theta_k) - V_i^2Y_{ii} \\\text{for} ~i = n_g+1,\dots n
    \end{split}
\end{cases}
\end{equation}
\par Linearizing (\ref{f1}) $-$ (\ref{g2}) around an operating point, with  $V_{i_0}$ as the voltage magnitude of bus $i$ at that point and defining a new variable $\nu_i = V_i/V_{i_0}$, we obtain
\begin{equation}
\label{sys_int}
\small
\begin{aligned}
\begin{split}
\arraycolsep=2pt\def\arraystretch{0.8} \left[\begin{array}{c}\Delta \dot{\boldsymbol{\delta}} \\ 
 \Delta \dot{\boldsymbol{\omega}} \\
 \Delta \dot{\boldsymbol{ E_q'}}\end{array}\right] =  &~\mathbf{M} \left[\begin{array}{c}\Delta \boldsymbol{\delta} \\ 
 \Delta \boldsymbol{\omega} \\
 \Delta \boldsymbol{ E_q'}\end{array}\right] +  ~\mathbf{N}\left[\begin{array}{c}\Delta \boldsymbol{\theta} \\ 
 \Delta \boldsymbol{\nu}
 \end{array}\right] + ~\mathbf{B} \left[\begin{array}{c}\Delta \boldsymbol{T_m} \\ 
 \Delta \boldsymbol{E_{fd}}
 \end{array}\right]
\\
\left[\begin{array}{c} \mathbf{0} \\ \mathbf{0}\end{array}\right] = &~\mathbf{C} \left[\begin{array}{c}\Delta \boldsymbol{\delta} \\ 
 \Delta \boldsymbol{\omega} \\
 \Delta \boldsymbol{ E_q'}\end{array}\right] +  ~\mathbf{D}\left[\begin{array}{c}\Delta \boldsymbol{\theta} \\ 
 \Delta \boldsymbol{\nu}
 \end{array}\right]
 \end{split}
\end{aligned}
\end{equation}
where, $\boldsymbol{\delta}$, $\boldsymbol{\omega}$, $\boldsymbol{E_{q}'}$, $\boldsymbol{\theta}$, and $\boldsymbol{\nu}$ are the vectorized state and algebraic variables of respective type, for instance, $\boldsymbol{\delta} =  \left[ \begin{array}{ccc}\delta_i & \dots & \delta_{n_g}\  \end{array} \right]^T$ and $\boldsymbol{\nu} = \left[ \begin{array}{ccc}\nu_i  & \dots & \nu_n \  \end{array} \right]^T$.  
Finally, eliminating the algebraic variables, we get
\begin{equation}\label{sys0}
\small
   \arraycolsep=2pt\def\arraystretch{0.8} \left[\begin{array}{c}\Delta \dot{\boldsymbol{\delta}} \\ 
 \Delta \dot{\boldsymbol{\omega}} \\
 \Delta \dot{\boldsymbol{ E_q'}}\end{array}\right] =  ~\mathbf{A} \left[\begin{array}{c}\Delta \boldsymbol{\delta} \\ 
 \Delta \boldsymbol{\omega} \\
 \Delta \boldsymbol{ E_q'}\end{array}\right]~+~\mathbf{B} \left[\begin{array}{c}\Delta \boldsymbol{T_m} \\ 
 \Delta \boldsymbol{E_{fd}}
 \end{array}\right]
\end{equation}
where, $\small {\mathbf{A}} =\small{ \mathbf{M - ND^{^{-1}}C}}$. It follows from the equations above that $\mathbf{A}$ is of the form
\begin{equation}
\small
\label{A}
    \mathbf{A} = \left[ \begin{array}{c c c} \mathbf{0} & \mathbf{I} & \mathbf{0}\\  \mathbf{A_{21}} & \mathbf{0} & \mathbf{A_{23}} \\ \mathbf{A_{31}} & \mathbf{0} & \mathbf{A_{33}} \end{array}\right]
\end{equation}
Note, in (\ref{A}), every $\mathbf{A_{ij}}$ block is a submatrix of $\mathbf{A}$ whose elements are derived later in the paper (see, Appendix B). Apart from the state variables $\Delta \delta_i$, $\Delta \omega_i$, and $\Delta E_{q_i}'$, the output variable $\Delta T_{e_i}$, which is the electromagnetic torque of generator $i$, is of specific interest to us. From the swing equation in (\ref{f2}) this is expressed as \textcolor{black}{$\Delta T_{e_i} = -\frac{2H_i}{\omega_s}~\Delta \dot \omega_i$}.

\par Following any disturbance in the system or perturbation in the inputs, the time-evolution of these state and output variables can be expressed as sum of damped sinusoids with modal frequencies $\omega_{d_r}$-s with differing amplitudes and phases. As a result, for each mode $r$, $\Delta T_{e_{i,r}}(t)$, $\Delta \delta_{i,r}(t)$ and $\Delta \omega_{i,r}(t)$ can be expressed as rotating phasors $-$ for details, see Appendix A. The notions of damping torque and damping power for a given mode originate from the phasor representation of $\Delta \vec T_{e_{i,r}}$ in the $\Delta \vec\delta_{i,r} - \Delta \vec\omega_{i,r}$ plane and the power resulting from the interaction of the torque and the speed. This is explained next. 

\section{Damping Power in a Multimachine System}
\label{secDP}

In any $i^{\text{th}}$ machine, for a mode $r$, let the average power of the electromagnetic torque $\Delta T_{e_{i,r}}(t)$ over a cycle \kc{starting from $t = t_0$} be denoted by $\dot W_{d_{i,r}}\kc{(t_0)}$, as shown below
\begin{equation}
\small
    \dot W_{d_{i,r}}\kc{(t_0)} = \frac{ \int_{t_0}^{t_0 + \frac{2\pi}{\omega_{d_r}}} \Delta T_{e_{i,r}}(t)~ \Delta \omega_{i,r}(t)~ dt }{\int_{t_0}^{t_0 + \frac{2\pi}{\omega_{d_r}}} dt}
\end{equation}
Using the phasor notation described in Appendix A, \textcolor{black}{let $\Delta \vec{T}_{e_{i,r}}(t) = \beta_1~e^{\sigma_rt}~\angle{\gamma_1}$ and $\Delta \vec{\omega}_{i,r}(t) = \beta_2~e^{\sigma_rt}~\angle{\gamma_2}$. Therefore, $\small{\dot W_{d_{i,r}}(t_0) =} $}
\kc{
\begin{equation}
\label{damp_power_1}
\small
\begin{aligned}
   = \frac{\omega_{d_r}}{2\pi}\int_{t_0}^{t_0 + \frac{2\pi}{\omega_{d_r}}} e^{2\sigma_rt}~ \beta_1  \cos(\omega_{d_r}t + \gamma_1)~ \beta_2 & \cos(\omega_{d_r}t + \gamma_2)~ dt \\
      = \frac{\beta_1~\beta_2~\omega_{d_r}}{4 \pi} \int_{t_0}^{t_0 + \frac{2\pi}{\omega_{d_r}}} e^{2\sigma_rt}~ \Big\{  \cos(\gamma_1 - \gamma_2)\\  + ~\cos(2\omega_{d_r}t~  & + ~ \gamma_1 + \gamma_2) \Big\}~dt\\
      = ~\frac{\beta_1~\beta_2~\omega_{d_r}}{4 \pi} ~\cos(\gamma_1 - \gamma_2)~ \frac{e^{2\sigma_rt_0}}{2~\sigma_r}~\Big\{ e^{ \frac{4\pi\sigma_r}{\omega_{d_r} }} &-  1 \Big\}\\
      + ~\frac{\beta_1~\beta_2~\omega_{d_r}}{4 \pi} \int_{t_0}^{t_0 + \frac{2\pi}{\omega_{d_r}}} e^{2\sigma_rt}~  \cos(2 & \omega_{d_r}t  + \gamma_1 + \gamma_2)~ dt ~
    \end{aligned}
\end{equation}
}
\textcolor{black}{Now, considering that our mode of interest is poorly-damped (as is the premise of our paper), implying $|\sigma_r|~ << \omega_{d_r}$, we may expand the exponential $e^{ \frac{4\pi\sigma_r}{\omega_{d_r} }}$ and neglect the second and higher order terms. On doing so, the first term in (\ref{damp_power_1}) reduces to}
\begin{equation*} \small
\begin{aligned}
 &\textcolor{black}{\frac{\beta_1~\beta_2~\omega_{d_r}}{4 \pi}~ e^{2\sigma_rt_0} ~\cos(\gamma_1 - \gamma_2)~ \frac{1 +\frac{4\pi\sigma_r}{\omega_{d_r}} - 1 }{2~\sigma_r}}\\ & \textcolor{black}{ 
 =\frac{1}{2} ~\beta_1~\beta_2~ e^{2\sigma_rt_0}~\cos(\gamma_1 - \gamma_2)}
    \end{aligned}
\end{equation*}
\textcolor{black}{With the same assumption that $\sigma_r$ is small, the second term in (\ref{damp_power_1}) becomes negligible, and can be ignored for mathematical tractability. This is because, with $|\sigma_r|~<<~\omega_{d_r}$, for a complete cycle of $\cos(2\omega_{d_r}t)$, the $e^{2\sigma_rt}$ term remains almost constant, and therefore, the positive and negative half cycles almost add to zero. 
Therefore, }
\begin{equation}
\label{damp_power}
\small
\begin{aligned}
   \textcolor{black}{ \dot W_{d_{i,r}}(t_0) }&\textcolor{black}{ ~\approx~ \frac{1}{2} ~\beta_1~\beta_2~ e^{2\sigma_rt_0}~\cos(\gamma_1 - \gamma_2)} \\ & \textcolor{black}{= \frac{1}{2} ~\Re \Big\{ \beta_1~ e^{\sigma_rt_0}~ \angle{\gamma_1}~~ \beta_2~ e^{\sigma_rt_0} ~\angle{-\gamma_2} \Big\} }\\ & \textcolor{black}{= \frac{1}{2} ~\Re \Big \{ \Delta \vec{T}_{e_{i,r}}(t_0) ~\Delta \vec{\omega}_{i,r}^*(t_0)\Big \}.}
    \end{aligned}
\end{equation}
\textcolor{black}{Hereafter, in the paper, assuming all phasors are computed at $t=t_0$ and powers are averaged over a cycle starting at $t_0$, we shall drop the argument $t_0$ from our expressions.}
\par 
From (\ref{damp_power}) it can be interpreted that $\dot W_{d_{i,r}}$ is the average power due to the component of $\Delta \vec{T}_{e_{i,r}}$ in the direction of $\Delta \vec{\omega}_{i,r}$. Let, the electromagnetic torque be decomposed as $\Delta \vec{T}_{e_{i,r}} = k_{d_{i,r}} \Delta \vec{\omega}_{i,r} + k_{s_{i,r}} ~(j\Delta \vec{\omega}_{i,r})$.
Substituting this in (\ref{damp_power}), we get
\begin{equation}
\small
    \label{kdwise}
 \dot W_{d_{i,r}} =  \frac{1}{2} \Re \Big \{ (k_{d_{i,r}} + j k_{s_{i,r}} )\Delta \vec{\omega}_{i,r}\Delta \vec{\omega}_{i,r}^*\Big \} = \frac{1}{2} k_{d_{i,r}} \big|\Delta \vec{\omega}_{i,r}\big|^2
\end{equation}
where, $k_{d_{i,r}} =\Re \{ \frac{\Delta \vec{T}_{e_{i,r}}}{\Delta \vec{\omega}_{i,r}}\} $ is the damping torque coefficient of machine $i$ for mode $r$. Therefore, we shall refer to $\dot W_{d_{i,r}}$ as the `average damping power' (or simply `damping power') of $\Delta T_{e_{i,r}}(t)$. 
For a system with $n_g$ machines, (\ref{damp_power}) can be extended as follows
\begin{equation}
\small
\label{Wr0}
    \dot W_{d_{r}} = \sum_{i=1}^{n_g}  \dot W_{d_{i,r}} = \frac{1}{2}~ \sum_{i=1}^{n_g}\Re \Big\{ \Delta\vec{T}_{e_{i,r}} ~\Delta\vec{\omega}_{i,r}^*  \Big\} 
\end{equation}
Next, we express this sum of damping powers in terms of system matrices by using the definition of electromagnetic torque (from the swing equation) and the linearized system description obtained in (\ref{sys0}) and (\ref{A}), as shown below.
\begin{equation}
\label{Temm}
\small
\begin{aligned}
    \Delta \boldsymbol{T_{e}}(s) &= -\frac{2~\mathbf{H}}{\omega_s}\Delta \boldsymbol{\dot \omega}(s) = -\frac{2~\mathbf{H}}{\omega_s}\Big \{ \mathbf{A_{21}}~\Delta\boldsymbol{\delta}(s) + \mathbf{A_{23}}~\Delta\boldsymbol{ E_q'}(s) \Big \}\\
    &= -\frac{2~\mathbf{H}}{\omega_s}~\bigg \{ \mathbf{A_{21}} + \mathbf{A_{23}}~(s\mathbf{I}-\mathbf{A_{33}})^{^{-1}}\mathbf{A_{31}}\bigg\}~\frac{\Delta\boldsymbol{\omega}(s)}{s} \\ &\overset{\Delta}{=}~ \mathbf{K}(s) ~\Delta\boldsymbol{\omega}(s)
    \end{aligned}
\end{equation}
Defining $\mathbf{K}_r = \mathbf{K}(j\omega_{d_r})$, we may re-write (\ref{Wr0}) as
\begin{equation}
\label{Wr1}
\small
    \begin{aligned}
      \dot W_{d_{r}} = \frac{1}{2} \Re \Big\{ \sum_{i=1}^{n_g} \sum_{j=1}^{n_g} \mathbf{K}_{ij,r} \Delta\vec{\omega}_{j,r}   \Delta\vec{\omega}_{i,r}^*  \Big\} =  \frac{1}{2} \Re \Big\{\Delta\vec{\boldsymbol \omega}_{r}^{^H} \mathbf{K}_{r} \Delta \vec{\boldsymbol \omega}_{r} \Big\} 
    \end{aligned}
\end{equation}
where, $\mathbf{K}_{ij,r}$ is the $(i,j)^{\text{th}}$ element of $\mathbf{K}_r$. We call $\dot W_{d_{r}}$ the `total damping power' of the system for mode $r$. 
\par Next, we simplify the expression in eqn (\ref{Wr1}) using the set of claims (1) $-$ (4) below. \text{Claims:}
\begin{equation*}
\small
    \begin{aligned}
      &(1)~ \mathbf{P}^{^{-1}}\mathbf{A}^{^T}_{\mathbf{33}}\mathbf{P} = \mathbf{A}_{\mathbf{33}}, \\ &(2) ~ \mathbf{A}^{^T}_{\mathbf{31}}\mathbf{P} = \frac{2~\mathbf{H}}{\omega_s}~\mathbf{A_{23}}, \\ 
      & (3)~  2~\mathbf{A}^{^T}_{\mathbf{21}}\mathbf{H} = 2~\mathbf{H}~\mathbf{A_{21}}\\
      &  (4) ~\forall~ \mathbf{x} \in \mathbb{C}^{n_g}, ~ \Re \{\mathbf{x}^{H} \mathbf{K}_{r} \mathbf{x} \} = \mathbf{x}^{H}\Re\{ \mathbf{K}_{r}\}~\mathbf{x} 
    \end{aligned}
\end{equation*}
where, $\mathbf{P}$ is a diagonal matrix of machine parameters with $\mathbf{P}(i,i) = \frac{T_{do_i}'}{x_{d_i}-x_{d_i}'}$. Claims (1) $-$ (3) are derived using the differential and algebraic equations of the system modeled in Section \ref{model}. These claims are then used to establish the symmetry of $\mathbf{K}_r$, which is in-turn used in proving claim (4). Detailed proof of these claims are outlined in Appendix B.

 \par Using claim (4) we further reduce eqn (\ref{Wr1}) as follows
\begin{equation}
\label{Wr2}
\small
    \begin{aligned}
      &\dot W_{d_{r}} =  \frac{1}{2} \Delta\vec{\boldsymbol \omega}_{r}^{^H} \Re \{\mathbf{K}_{r} \} \Delta \vec{\boldsymbol \omega}_{r} 
    \end{aligned}\vspace{-0.2cm}
\end{equation}
where, $\small{ \Re \{\mathbf{K}_{r} \} =} $
\begin{equation*}\label{reK}
\small
   \begin{aligned}
       &= -\Re \bigg\{ \frac{2~\mathbf{H}}{j\omega_{d_r}~\omega_s}~\bigg(\mathbf{A_{21}} + \mathbf{A_{23}}~(j\omega_{d_r}\mathbf{I}-\mathbf{A_{33}})^{^{-1}}\mathbf{A_{31}}\bigg)\bigg\}\\
     &\begin{split}=-\frac{2~\mathbf{H}}{\omega_s}~\mathbf{A_{23}}~\Re \bigg\{(-j\omega_{d_r}\mathbf{I}-\mathbf{A_{33}})(-j\omega_{d_r}\mathbf{I}-\mathbf{A_{33}})^{^{-1}} \\\frac{(j\omega_{d_r}\mathbf{I}-\mathbf{A_{33}})^{^{-1}}}{j\omega_{d_r}}\bigg\}~\mathbf{A_{31}}\end{split}\\
     &= -\frac{2~\mathbf{H}}{\omega_s}~\mathbf{A_{23}}~\Re \bigg\{\frac{(-j\omega_{d_r}\mathbf{I}-\mathbf{A_{33}})(\omega_{d_r}^2\mathbf{I}+\mathbf{A_{33}}^2)^{^{-1}}}{j\omega_{d_r}}\bigg\}~\mathbf{A_{31}}\\
    &= \frac{2~\mathbf{H}}{\omega_s}~\mathbf{A_{23}}~(\omega_{d_r}^2\mathbf{I}+\mathbf{A_{33}}^2)^{^{-1}}\mathbf{A_{31}} 
    \end{aligned}
\end{equation*}
This, along with claim (2) when substituted in eqn (\ref{Wr2}) gives
\begin{equation}
\label{Wr3}
\small
    \begin{aligned}
      &\dot W_{d_{r}} =  \frac{1}{2}~ \Delta\vec{\boldsymbol \omega}_{r}^{^H}~\mathbf{A}^{^T}_{\mathbf{31}}~\mathbf{P} ~(\omega_{d_r}^2\mathbf{I}+\mathbf{A_{33}}^2)^{^{-1}}\mathbf{A_{31}}~ \Delta \vec{\boldsymbol \omega}_{r}
    \end{aligned}    
\end{equation}

\section{Consistency of Damping Power with Power Dissipation in Synchronous Machine Windings}
\label{field}

Considering the third-order system model in Section \ref{model}, the only source of power dissipation in the machines is in the field windings. For any mode $r$, we denote the average power dissipation in the winding of machine $i$ by $\dot W_{f_{i,r}}$. Following the phasor notation as discussed, this is expressed as
\begin{equation}
\small
    \label{wfi}
    \begin{aligned}
     \dot W_{f_{i,r}} &=  \frac{1}{2} \Re \Big\{ (\Delta\vec{I}_{f_{i,r}}R_{f_i}) ~\Delta\vec{I}_{f_{i,r}}^*  \Big\} = \frac{1}{2}~R_{f_i}~\Big|\Delta\vec{I}_{f_{i,r}}\Big|^2\\
     &= \frac{1}{2} ~\frac{T_{do_i}'}{x_{d_i}-x_{d_i}'} \Big|\Delta \vec{\dot{ E}}_{q_{i,r}}'\Big |^2
    \end{aligned}
\end{equation}
where, $\Delta \vec{I}_{f_{i,r}}$ and $\Delta \vec{\dot{ E}}_{q_{i,r}}'$ are the phasors of the field current and derivative of transient e.m.f. due to field flux linkage, respectively. Further, using notations from (\ref{phasdef_2}) (Appendix A) we may write 
\begin{equation}
    \small
    \label{eqdot}
    \textcolor{black}{\Delta \vec{\dot{ E}}_{q_{i,r}}' = j \omega_{d_r}~\Delta \vec{ E}_{q_{i,r}}' = j \omega_{d_r}~2~c_r~ e^{\sigma_rt_0} ~\psi_{E^'_{q_{i,r}}}} ~.
\end{equation}
\kc{We define, $\hat{c}_r~=~c_r~e^{\sigma_rt_0}$}. Next, substituting \eqref{eqdot} in (\ref{wfi}) we get
\begin{equation}
    \small
    \label{wfi2}
     \dot W_{f_{i,r}}
     = 2 ~\kc{|\hat{c}_r|^2} ~ \frac{T_{do_i}'}{x_{d_i}-x_{d_i}'}~\omega_{d_r}^2 ~ \Big|\psi_{E^'_{q_{i,r}}}\Big |^2 ~.
\end{equation}
We obtain the total power dissipation for the mode by adding the dissipation in individual machines. This is shown below. 
\begin{equation}
    \small
    \label{wf}
    \dot W_{f_{r}} = \sum_{i=1}^{n_g} \dot W_{f_{i,r}}
     = 2  ~\kc{|\hat{c}_r|^2} ~ \omega_{d_r}^2 ~ \Psi^H_{E^'_{q_{r}}}~\mathbf{P}~\Psi_{E^'_{q_{r}}}
     \vspace{-0.2cm}
\end{equation}
where, $\small {\Psi_{E^'_{q_{r}}} =  \left[ \begin{array}{ccccc}\psi_{E^'_{q_{1,r}}} & \dots &\psi_{E^'_{q_{i,r}}} &\dots &\psi_{E^'_{q_{{n_g},r}}}  \end{array} \right]^T}$.
\par To show that our notion of total damping power, as defined in Section \ref{secDP}, is consistent with the power dissipated in the system, we need to prove that for any mode $r$, 
$\dot W_{f_{r}}$ is equal to $\dot W_{d_{r}}$. 
To do so, we make the following algebraic manipulations.
\par First, since $\lambda_r$ is an eigenvalue of the system, we may write $\mathbf{A}\Psi_r = \lambda_r\Psi_r$, where, the right eigenvector $\Psi_r = \left[ \begin{array}{ccc} \Psi_{\delta_{r}} & \Psi_{\omega_{r}}  &\Psi_{E^'_{q_{r}}} \end{array} \right]^T$. Next, using the structure of matrix $\mathbf{A}$ as in (\ref{A}), we can split this into the following equations
\begin{equation}\label{Psidef}
\small
    \begin{aligned}
      \Psi_{E^'_{q_{r}}} = (\lambda_r\mathbf{I} - \mathbf{A_{33}})^{-1}~\mathbf{A_{31}}~\Psi_{\delta_{r}} ~~~\text{and} ~~~
      \Psi_{\delta_{r}} = \frac{1}{\lambda_r} \Psi_{\omega_{r}}
    \end{aligned}
\end{equation}
Using these, along with (\ref{phasdef}) describing $\Delta \vec{\boldsymbol{\omega}}_r = 2~\kc{\hat{c}_r}~\Psi_{\omega_{r}}$, we may re-write (\ref{wf}) as follows
\begin{equation}
    \small
    \label{wf2}
    \begin{aligned}
      \dot W_{f_{r}} = 2 ~|c_r|^2 ~\omega_{d_r}^2~ \frac{\Delta \vec{\boldsymbol{\omega}}_r^H}{2~c_r^*~\lambda_r^*}&~\mathbf{A}^{^T}_{\mathbf{31}}~(\lambda_r^*\mathbf{I} - \mathbf{A_{33}}^T)^{-1}~\mathbf{P}\\&~(\lambda_r\mathbf{I} - \mathbf{A_{33}})^{-1}~\mathbf{A_{31}}~ \frac{\Delta \vec{\boldsymbol{\omega}}_r}{2~c_r~\lambda_r} .
    \end{aligned}
\end{equation}
Finally, considering that our mode of interest is poorly-damped $|\sigma_r|~<< \omega_{d_r}$ -- as is the premise of \cite{tsingua1}, and for consistency this paper's too, we substitute $\lambda_r = j\omega_{d_r}$ in \eqref{wf2}. This, along with use of claim (1) reduces \eqref{wf2} as follows
\begin{equation}
    \small
    \begin{aligned}
    \label{wf_total}
      \dot W_{f_{r}} = \frac{1}{2} ~ \Delta \vec{\boldsymbol{\omega}}_r^H ~ \mathbf{A}^{^T}_{\mathbf{31}}~ (-j\omega_{d_r}\mathbf{I} &- \mathbf{P}\mathbf{A_{33}}\mathbf{P}^{-1})^{^{-1}}~\mathbf{P}\\&~(j\omega_{d_r}\mathbf{I} - \mathbf{A_{33}})^{^{-1}}~\mathbf{A_{31}}~\Delta \vec{\boldsymbol{\omega}}_r\\
      =  \frac{1}{2}~ \Delta\vec{\boldsymbol \omega}_{r}^{^H}~\mathbf{A}^{^T}_{\mathbf{31}}~\mathbf{P} ~(\omega_{d_r}^2&\mathbf{I}+\mathbf{A_{33}}^2)^{^{-1}}\mathbf{A_{31}}~ \Delta \vec{\boldsymbol \omega}_{r} ~= ~\dot W_{d_r} .
    \end{aligned}
\end{equation}
Thus, for any given mode, the equivalence of the total damping power and the sum of average rate of change of energy dissipations in the machine windings is established. 
\par \textit{Remarks:} (1) While the concepts of damping torque and damping power are derived using the linearlized system models, the average rate of energy dissipation in windings are more fundamental and does not limit itself to small-signal analysis.
\par (2) For machines with detailed models, the power dissipation in damper windings should be added to $\dot W_{f_{r}}$ to obtain the total dissipation in the system. \textcolor{black}{Average power dissipation in the damper winding is expressed as \cite{chen2}}
\begin{equation}
    \small
    \label{wg_i}
     \textcolor{black}{\dot W_{g_{i,r}}
     = 2  ~\kc{|\hat{c}_r|^2} ~\frac{T_{qo_i}'}{x_{q_i}-x_{q_i}'}~\omega_{d_r}^2~ \Big|\psi_{E^'_{d_{i,r}}}\Big |^2 } .
\end{equation}
$E^'_{d_{i,r}}$ is the state variable describing the dynamics of the damper winding transient e.m.f. Further, with dynamics of the excitation systems modeled, the  expression of $\dot W_{f_{r}}$ in (\ref{wf}) would get modified to include the effect of $\Delta \boldsymbol{E_{fd}}$ as 
\begin{equation}
\begin{aligned}
    \small
    \label{wfex_i}
    \textcolor{black}{ \dot W_{f_{i,r}}
     = 2} ~& \textcolor{black}{ ~\kc{|\hat{c}_r|^2} ~\bigg\{ \frac{T_{do_i}'}{x_{d_i}-x_{d_i}'}~\omega_{d_r}^2~ \Big|\psi_{E^'_{q_{i,r}}}\Big |^2} \\&\textcolor{black}{- \frac{1}{x_{d_i}-x_{d_i}'}~\Re \Big( j\omega_{d_r}~\psi_{E^'_{q_{i,r}}}~\psi^{*}_{E_{fd_{i,r}}} \Big) \bigg\} } .
     \end{aligned}
\end{equation}
However, the expression for $\dot W_{d_{r}}$ in (\ref{Wr1}) would remain the same (with block matrices $\mathbf{A_{23}}$, $\mathbf{A_{31}}$, and $\mathbf{A_{33}}$ larger in dimensions to account for the additional state variables like $\Delta \boldsymbol{ E_d'}$, $\Delta \boldsymbol{E_{fd}}$, etc. now concatenated to the vector $\Delta \boldsymbol{ E_q'}$) and the equality of total damping power and total 
power dissipation would still be true. This will be demonstrated in Section \ref{result}. 

\par
\begin{figure}[H]
    \centering
    \vspace{-0.2cm}
    \hspace{-0.2cm}
    \includegraphics[width = 1.02\linewidth]{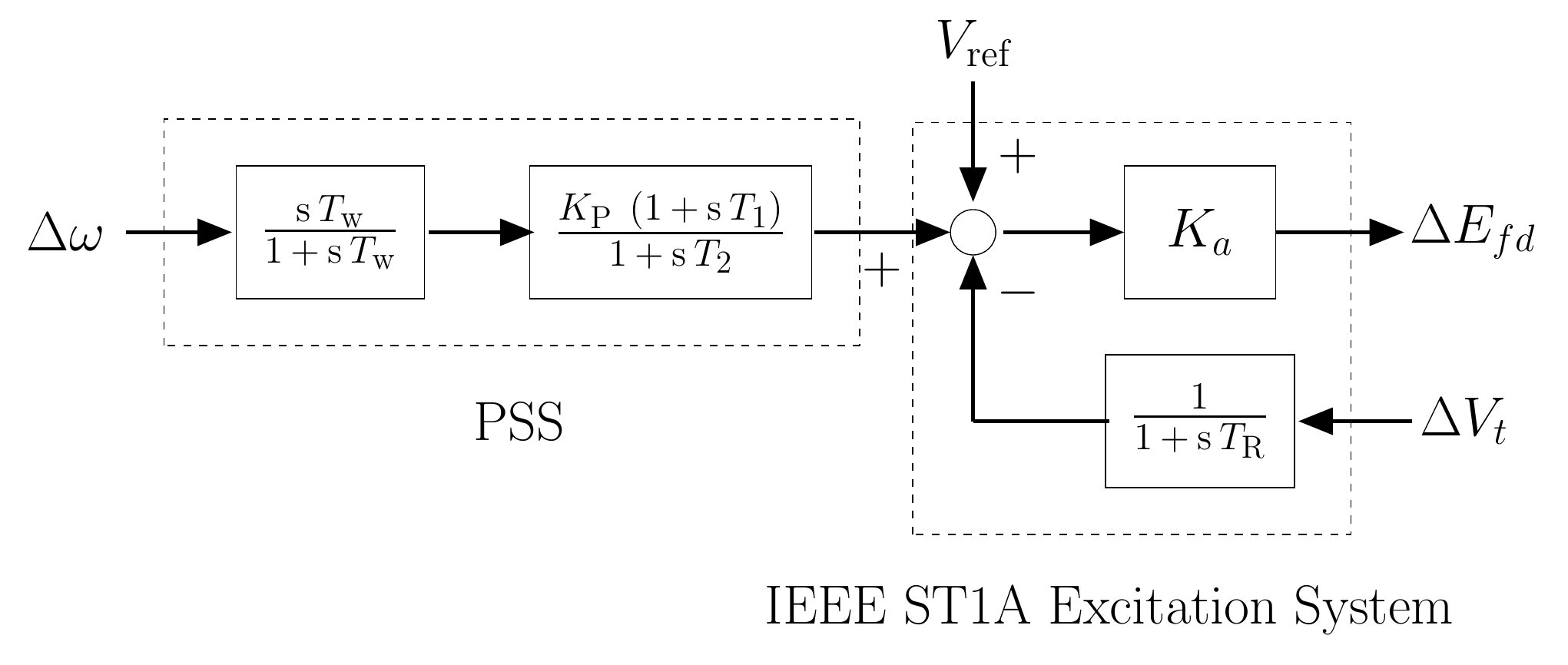}
    \caption{\textcolor{black}{PSS and IEEE ST$1$A excitation system.}}
    \label{pss}
\end{figure}

 \textcolor{black}{(3) Additionally, with a power system stabilizer (PSS) and an IEEE ST1A exciter modeled, as shown in Fig. \ref{pss}, three new states will be added $-$ one each due to the washout block, the lead-lag compensator, and the time constant of the transducer. Concatenating these to the existing state variables, the block matrices $\mathbf{A_{23}}$, $\mathbf{A_{31}}$, and $\mathbf{A_{33}}$ would be further expanded. Additionally, due to the speed feedback to the washout block, $\mathbf{A_{32}}$ term would now be non-zero, implying, $\mathbf{K}_{r} = \frac{2~\mathbf{H}}{\omega_s}~\bigg(\frac{\mathbf{A_{21}}}{j\omega_{d_r}} + \mathbf{A_{23}}~(j\omega_{d_r}\mathbf{I}-\mathbf{A_{33}})^{^{-1}}(\frac{\mathbf{A_{31}}}{j\omega_{d_r}}~+~\mathbf{A_{32}})\bigg) $. The equality of total damping power $\dot W_{d_{r}}$ (as calculated with the modified $\mathbf{K}_{r}$) and that of the sum of $\dot W_{f_{r}}$ and $\dot W_{g_{r}}$ will be demonstrated in Section \ref{result}.}
\vspace{-0.2cm}
\par (4) Next, let us explore the importance of our deductions in the context of stability monitoring. From \cite{chen2}, we know that, the relative damping contribution of individual generators can be inferred from the power dissipations in their windings. Further, our paper establishes $\sum \dot W_{d_{i,r}} = \sum (\dot W_{f_{i,r}} + \dot W_{g_{i,r}})$, which building on derivations in \cite{chen2} leads to the claim that the stability margin of the $r^{th}$ mode $\sigma_r \propto -\sum \dot W_{d_{i,r}}$. Note that, estimating the variables $\Delta T_{e_i}$ and $\Delta \omega_i$ are relatively simpler. Because, either they can be directly measured, as with rotor speed, or estimated from measurable outputs, like torque from power. And therefore, damping power $\dot  W_{d_{i,r}}$ of individual machines can be easily calculated from their terminal measurements upon filtering for the mode of interest. This has the potential for future monitoring applications, like determining stability margin of a mode by measuring total damping power contribution from all generators. \kcc{To this end, it is important to clarify that, we do not claim efficacy over the existing mode-metering algorithms like \cite{trud_modemeter} (which would anyways be required to identify the poorly-damped modes in the system), instead, we propose $\sum \dot W_{d_{i,r}}$ as a complementary measure of stability margin for the mode, with individual $\dot  W_{d_{i,r}}$-s as indices for identifying the prospective generator locations for damping enhancement.  }
\vspace{-0.11cm}
\section{Distribution Factors: Expressing 
Damping Power of Each Machine as Weighted Sum of Dissipations in All Machines}
\label{distri}
 With the equality in (\ref{wf_total}) now proved, it might be tempting to 
draw an intuitive conclusion that such a power balance should hold for individual generators. However, this is not the case. This is because, 
the modeshape of  $E^'_q$
of each generator is a function of modeshapes of speed-deviation of all other machines and vice-versa (see, (\ref{Psidef})). Therefore, the mathematical representation of damping power in each machine is in reality an abstraction of dissipative effects coming from the windings of all machines in the system including it's own.
To that end, we next derive the distribution factors describing the fractional contribution of windings of different machines
in constituting the damping power of a single machine. 

Considering the system model described in Section \ref{model}, in this section, we express the damping power of each generator $\dot W_{d_{i,r}}$ as a linear combination of individual $\dot W_{f_{i,r}}$-s. Recall, from (\ref{kdwise})
\begin{equation} \small
    \label{wfi3}
    \begin{aligned}
   \dot W_{d_{i,r}} = \frac{1}{2}~ k_{d_{i,r}}~ \Big|\Delta \vec{\omega}_{i,r}\Big|^2  = 2  ~\kc{|\hat{c}_r|^2} ~k_{d_{i,r}}~ \Psi^H_{\omega_{r}}~\mathbf{I}_i~\Psi_{\omega_{r}}
   \end{aligned}
\end{equation}
where, $\mathbf{I}_i$ is a $n_g-$dimensional square matrix with the $(i,i)^{\text{th}}$ entry as $1$ and remaining all entries as zeros. Next, from the eigen decomposition of $\mathbf{A}$ as before, we may write $\Psi_{\omega_{r}} = \lambda_r~(\lambda_r^2~\mathbf{I} ~-~ \mathbf{A_{21}})^{-1}~\mathbf{A_{23}} \overset{\Delta}{=} \mathbf{Q}~\Psi_{E^'_{q_{r}}}$. Substituting this, (\ref{wfi3}) may be expressed as
\begin{equation} \small \label{wfi4}
    \begin{aligned}
    \dot W_{d_{i,r}} &=  2  ~\kc{|\hat{c}_r|^2} ~k_{d_{i,r}}~ \Psi^H_{E^'_{q_{r}}}~\mathbf{Q}^H~\mathbf{I}_i~\mathbf{Q}~\Psi_{E^'_{q_{r}}}\\
    &= 2 ~\kc{|\hat{c}_r|^2} ~k_{d_{i,r}}~ \sum_{\ell = 1}^{n_g}~\sum_{j=1}^{n_g}~Q_{i\ell}^*~Q_{ij}~\psi^*_{E^'_{q_{\ell,r}}}~\psi_{E^'_{q_{j,r}}}\\
    &= \sum_{j=1}^{n_g}~2 ~\kc{|\hat{c}_r|^2} ~k_{d_{i,r}}~ \frac{Q_{ij}~\sum_{\ell = 1}^{n_g}~Q_{i\ell}^*~\psi^*_{E^'_{q_{\ell,r}}}}{\psi^*_{E^'_{q_{j,r}}}} ~\Big|\psi_{E^'_{q_{j,r}}}\Big|^2 \\
    &= \sum_{j=1}^{n_g}~\frac{k_{d_{i,r}}}{P_{j}~\omega_{d_r}^2}~ \frac{Q_{ij}~\psi^*_{\omega_{i,r}}}{\psi^*_{E^'_{q_{j,r}}}}~\dot W_{f_{j,r}}
    \end{aligned}
\end{equation}
 where, $P_j$ is the diagonal element $\mathbf{P}(j,j) = \frac{T_{do_j}'}{x_{d_j}-x_{d_j}'}$. Next, we define $\beta_{ij}$ as the component of $\Delta \vec{\omega}_{i,r}$  due to $\Delta \vec{E'}_{q_{j,r}}$  
 \begin{equation} \small
     \label{beta}
     \beta_{ij} \overset{\Delta}{=} \Big(~\frac{Q_{ij}~\Delta \vec{E'}_{q_{j,r}}}{\Delta\vec{\omega}_{i,r}}~\Big) ^{*} = \frac{Q_{ij}^*~\psi^*_{E^'_{q_{j,r}}}}{\psi^*_{\omega_{i,r}}} .
 \end{equation}
 Since, $\psi^*_{\omega_{i,r}}= \sum_{j = 1}^{n_g}~Q_{ij}^*~\psi^*_{E^'_{q_{j,r}}}$, we may say $Q_{ij}^*~\psi^*_{E^'_{q_{j,r}}}$ is the contribution of $\psi^*_{E^'_{q_{j,r}}}$  in the modeshape $\psi^*_{\omega_{i,r}}$ . Therefore,
 \begin{equation}
 \label{alpha0}
     \small
     \dot W_{d_{i,r}} = \sum_{j=1}^{n_g}~\frac{k_{d_{i,r}}~|Q_{ij}|^2}{P_{j}~\omega_{d_r}^2~\beta_{ij}}~\dot W_{f_{j,r}} 
 \end{equation}
 Since, $\dot W_{d_{i,r}}$-s and $\dot W_{f_{i,r}}$-s are real quantities, the imaginary part of (\ref{alpha0}) is zero. 
 Hence,
 \begin{equation}
 \label{alpha1}
     \small
     \dot W_{d_{i,r}} = \sum_{j=1}^{n_g}~\frac{k_{d_{i,r}}~|Q_{ij}|^2}{P_{j}~\omega_{d_r}^2~}~\Re\big(\frac{1}{\beta_{ij}}\big)\dot W_{f_{j,r}} \overset{\Delta}{=} \sum_{j=1}^{n_g} \alpha_{ij}~\dot W_{f_{j,r}} 
 \end{equation}
 We call $\alpha_{ij}$-s the `distribution factors,' because, for a fixed $i$, the ratios $\alpha_{ij} ~\frac{\dot W_{f_{j,r}}}{\dot W_{d_{i,r}}}$ for $j = 1$ to $n_g$, are the \textcolor{black}{fractions in which the damping power of generator $i$ is derived from the power dissipation in the windings of the machines $1$ to $n_g$. Further, looking from the other side, fixing a machine $j$, the factors $\alpha_{ij}$-s describe the ratios in which the power dissipation in that machine winding is \textit{distributed} in the `abstract' damping power of all other machines.}
\vspace{-0.3cm}
\subsection{\kcc{Connection to the Heffron-Phillips Model}} 
\kcc{We know from the Heffron-Phillips model \cite{kundur} of a SMIB system that, for a third-order machine model ($K1-K4$ model), the angle between the $\Delta \vec{T}_{e_{i,r}}$ and $\Delta \vec{\omega}_{i,r}$ phasors is determined by the field circuit time-constant. Higher the resistance of the field circuit, smaller is the angle, and therefore, higher is the damping power of the generator. This is consistent with the results in \cite{tsingua1} that for a SMIB system, the damping power of the machine is derived exclusively from the power dissipation in its field circuit. Extending the same to the Heffron-Phillips model for multimachine systems, we see (from Fig. 6 in \cite{chow_pss}) that the damping power of each generator has contributions from the field circuit dissipations in multiple other machines in the system -- depending on the relative participation of those machines in the mode of interest. However, given the complexity in calculating the modewise $K1-K4$ constants in a multimachine system, we, through the distribution factors derived in this section, offer an alternative path to express the damping powers of each generator as a weighted sum of field winding dissipations of all machines in the system, including it's own.  }
\vspace{-0.2cm}
\subsection{\kcc{Potential Application in Understanding Dissipative Contribution from PSS and Other Controllers }}
\textcolor{black}{
\kcc{The lead-lag compensators in a $\Delta \omega$-PSS are designed knowing the angular relationship between the $\Delta \vec{T}_{e_{i,r}}$ and $\Delta \vec{\omega}_{i,r}$ phasors, and the phase-shift required to rotate $\Delta \vec{T}_{e_{i,r}}$ further in the direction of $\Delta \vec{\omega}_{i,r}$. This phase-shift introduces additional damping for the mode \kcc{by increasing} the \kcc{total} power loss in the system for the mode for which the PSS is designed. However, since a PSS in one location might negatively affect the damping in some other location, dissipation in some individual generators may be reduced.} 
} 
\kcc{While, methods like the one in \cite{chow_pss} quantify the effect of PSS on damping powers of individual machines, they do not describe how the dissipations in their windings would change. 
To this end, the distribution factors connecting damping and dissipative powers become useful. Also, this is not limited to PSSs in generators, if derived for higher-order models, the analysis can be extended to other types of controllers.}


\par \textcolor{black}{While  distribution factors may not have direct usefulness in terms of monitoring or controlling a mode, we believe they serve as an important tool that bridge an insightful link between two apparently different frameworks.
}

\section{Case Studies}
\label{result}
We now verify the aforementioned claims on the fundamental frequency phasor models of IEEE $4$-machine \cite{kundur} and $16$-machine \cite{nilthesis} test systems. In each case we consider two types of models: (a) \textit{Simplified model} with assumptions described in Section \ref{model}, and (b) \textit{Detailed model} considering $4$th-order synchronous machine dynamics (including $1$ damper winding) along with exciters, where the network is still assumed to be lossless and loads are of constant power nature.
\vspace{-0.1cm}
\subsection{IEEE $2-$area $4-$machine Kundur Test System}
\begin{figure}[h]
    \centering
    \includegraphics[width = 0.95\linewidth]{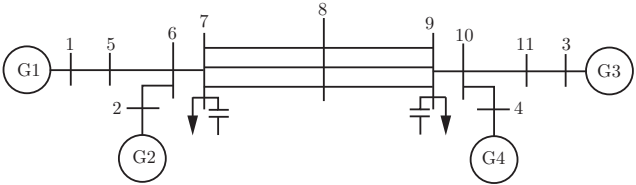}
    \caption{Single-line diagram of $2-$area $4-$machine test system.}
    \label{fig_4mc}
\end{figure}

Consider 
the $4-$machine system \cite{kundur} shown in Fig. \ref{fig_4mc} with a total load of $2,734$ MW under nominal condition. 
\par
\begin{figure}
\begin{subfigure}{\linewidth}
\centering
\includegraphics[width = \linewidth]{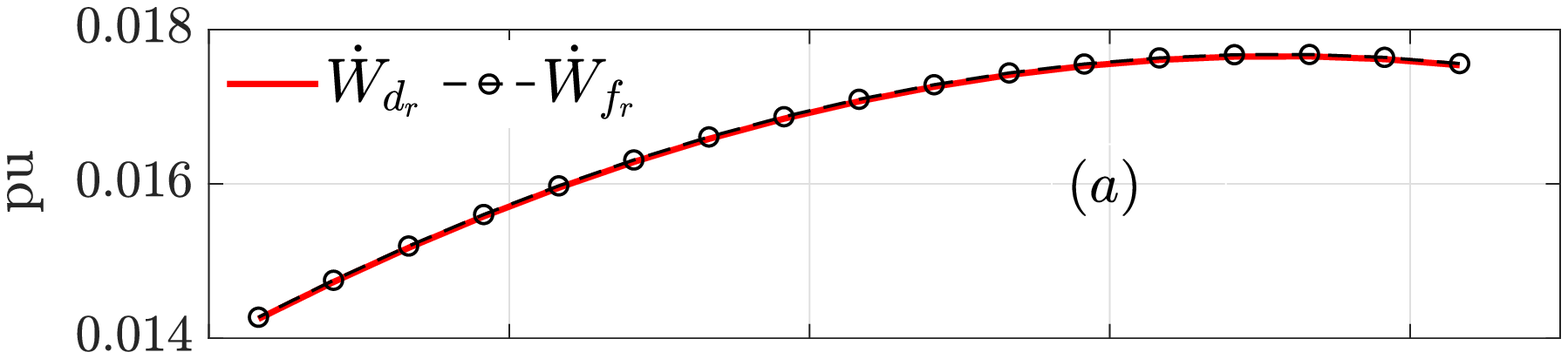}
\end{subfigure}
\begin{subfigure}{\linewidth}
\centering
\vspace{-0.12cm}
\includegraphics[width = \linewidth]{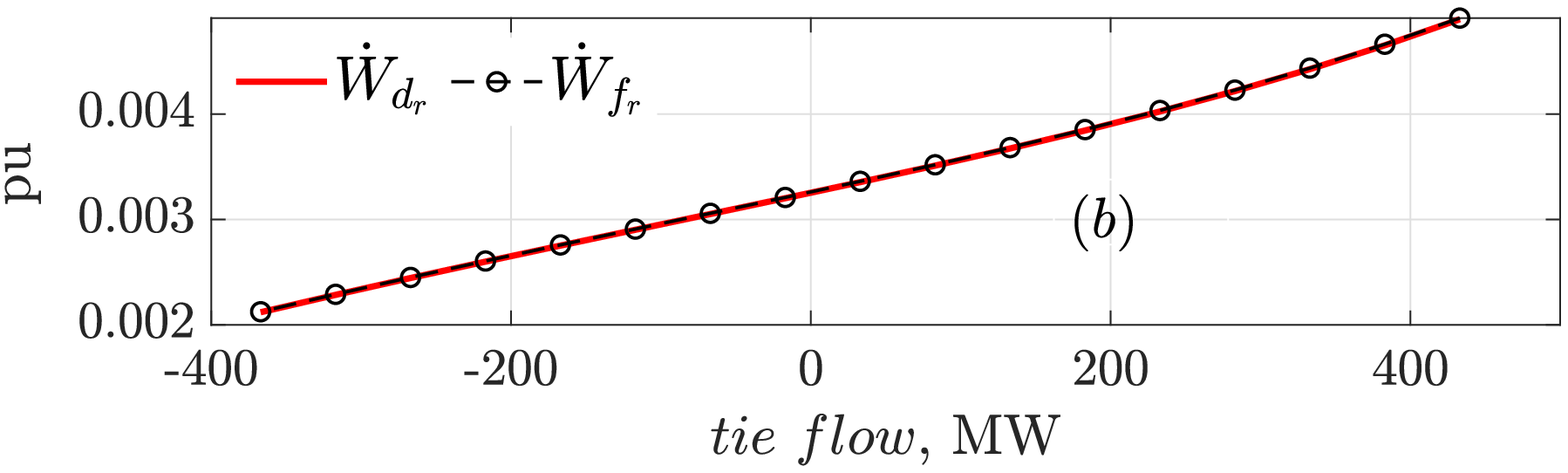}
\end{subfigure}
    \caption{Equality of total damping power with sum of average power dissipation in field windings of all generators across different operating points in simplified $4-$machine system model 
    for (a) $0.69$ Hz and (b) $1.04$ Hz modes.}
    \label{consis_4mc_1}
\end{figure}

 \textit{(a) Simplified model:} 
Under nominal loading, there are three poorly-damped modes, see Table I. 
Given modeling assumptions, the only source of damping is in the field windings of the generators. Therefore, following our proposition, we need to show that for small perturbations in the system, for each of these oscillatory modes, the sum of average damping powers is numerically equal to the sum of power dissipations in the field windings, across the operating points. This is demonstrated in the Figs. \ref{consis_4mc_1} (a) and (b) for two of the three modes. The operating point is varied by progressively reducing the tie flow between buses $7$ and $9$ from $433$ MW under nominal condition to $-400$ MW while maintaining the total load of the system constant.
\begin{figure}
  \begin{subfigure}{.5\columnwidth}
    \centering
    \includegraphics[width=1.1\linewidth]{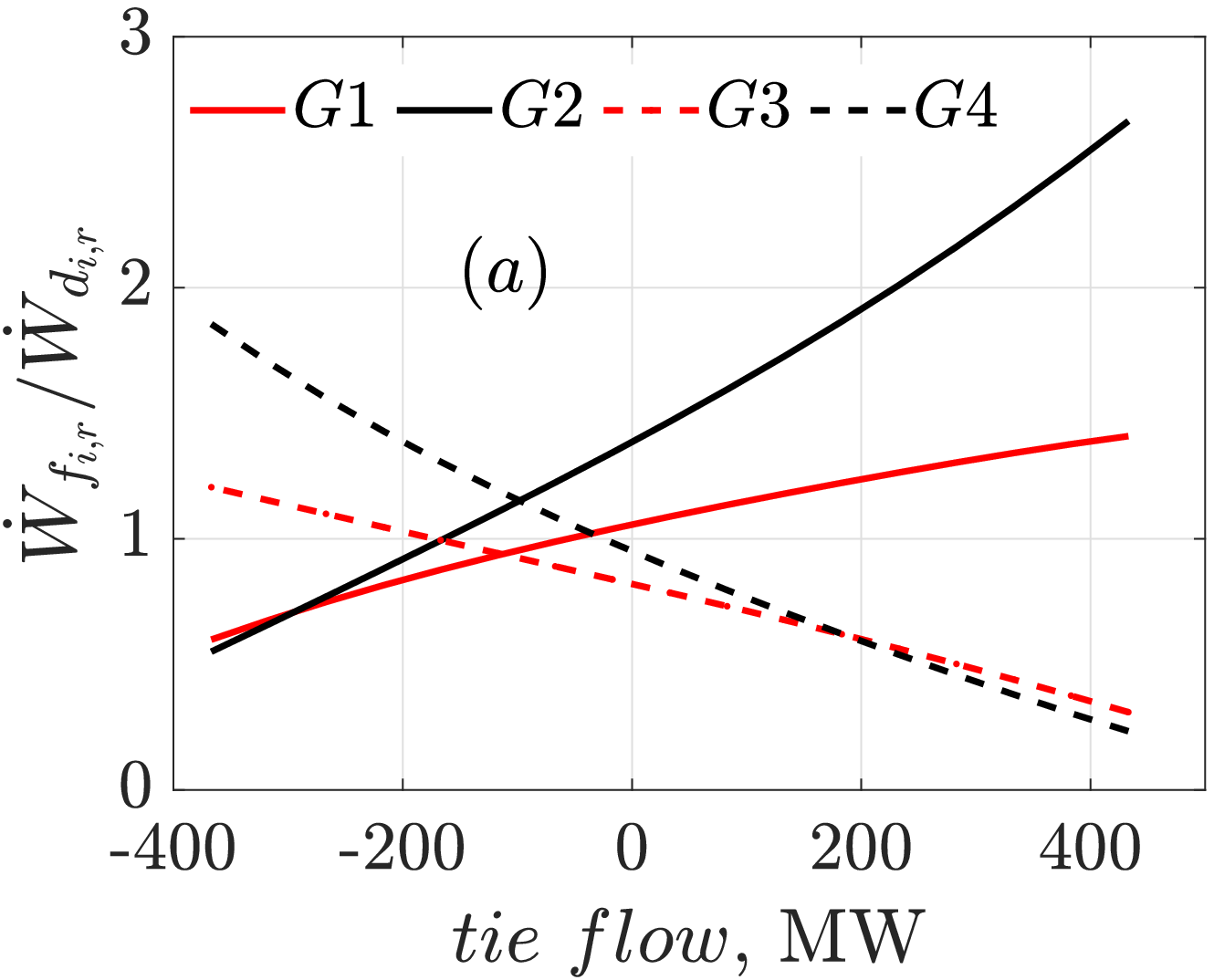}
  \end{subfigure}%
  \begin{subfigure}{.5\columnwidth}
    \centering
    \includegraphics[width=1.1\linewidth]{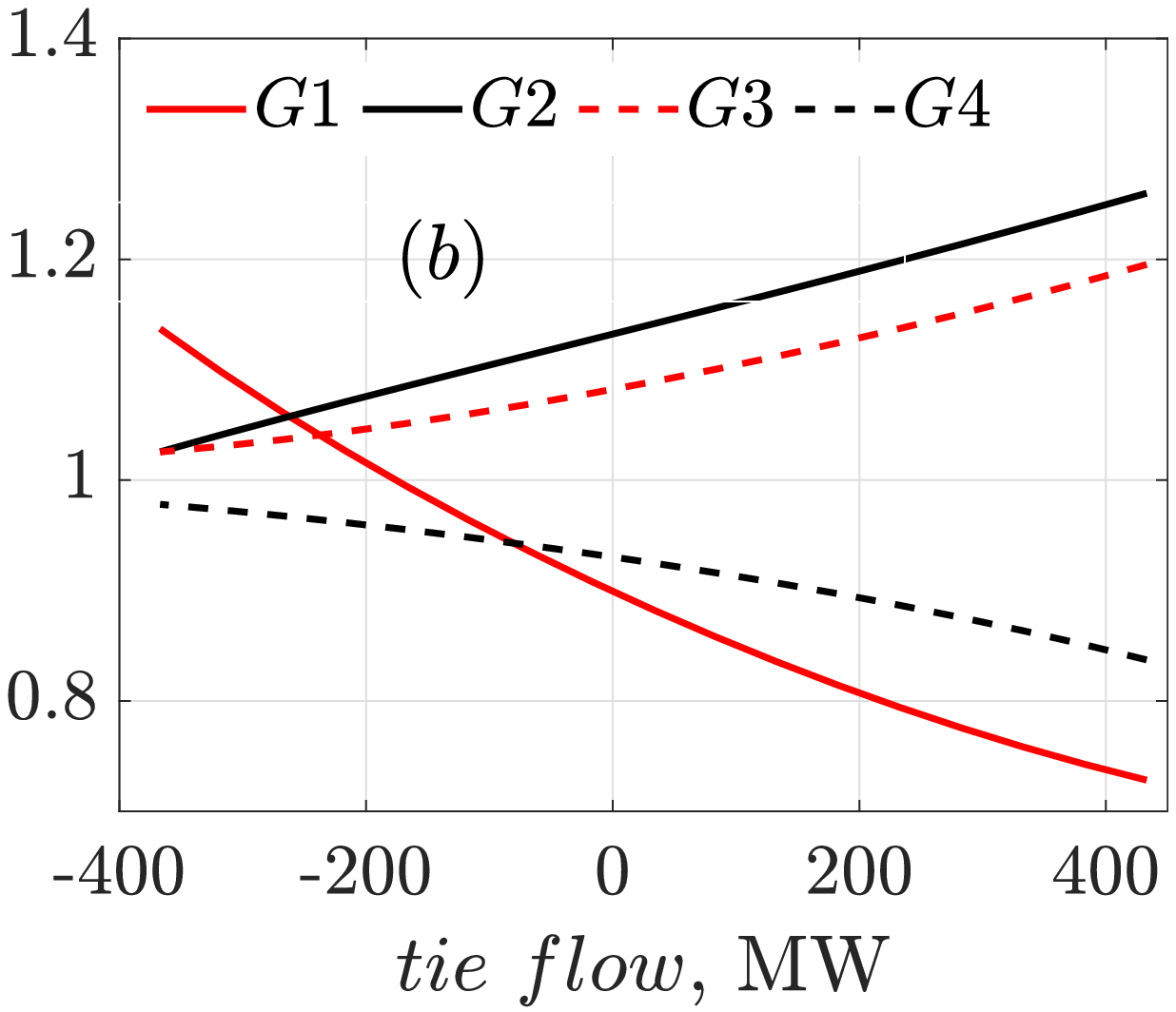}
  \end{subfigure}%
  \caption{Ratios of average power dissipation in field windings ($\dot W_{f_{i,r}}$) and that due to damping torques ($\dot W_{d_{i,r}}$) for individual machines across different operating points in simplified $4-$machine system model 
  for (a) $0.69$ Hz and (b) $1.04$ Hz modes.}
  \label{ratio_4mc_1}
  \vspace{-0.4cm}
  \end{figure}
\par We now validate our claim that although the total power dissipation is equal to the total damping power at the system level, this is not necessarily true for individual machines.  
In Figs \ref{ratio_4mc_1} (a) and (b), the ratios of dissipation power to damping power is plotted for each of the $4$ machines for the $0.69$ Hz and $1.04$ Hz modes, respectively. Observe that the ratios show strict monotonicity with change in operating points and under no circumstance they are equal to $1$ all at once.
\begin{table}
\centering
\setlength{\tabcolsep}{3pt}
\renewcommand{\arraystretch}{1.5}
\caption{\textcolor{black}{POORLY-DAMPED MODES IN IEEE 4-MACHINE SYSTEM}}
\begin{tabular}{c||c||c|c}
\hline
Machine Model & \begin{tabular}[c]{@{}c@{}}Eigenvalues\\ $\lambda_r  = \sigma_r \pm j\omega_{d_r}$\end{tabular} & \begin{tabular}[c]{@{}c@{}}Modal freq.\\ $f_r$ (Hz)\end{tabular} & \begin{tabular}[c]{@{}c@{}}Damp. ratio\\ $\zeta_r$\end{tabular} \\ \hline \hline
\multirow{3}{*}{Simplified model} & $-0.1183 \pm j 4.3816$ & $0.69$ & $0.027$ \\& $-0.1444 \pm j 6.2779$ & $1.00$ & $0.023$ \\
 & $-0.1544 \pm j 6.5330$ & $1.04$ & $0.024$ \\ \hline 
\begin{tabular}[c]{@{}c@{}}Detailed model \\(with DC$1$A exciters)\end{tabular} & $-0.1231 \pm j 4.2130$ & $0.67$ & $0.029$ \\ \hline
\begin{tabular}[c]{@{}c@{}}Detailed model (with \\ ST$1$A exciters \& PSS)\end{tabular} & $-0.1477 \pm j 4.8649$ & $0.77$ & $0.030$ \\ \hline
\end{tabular}
\end{table}
\par
\begin{figure}[h]
\vspace{-0.2cm}
      \includegraphics[width=\linewidth]{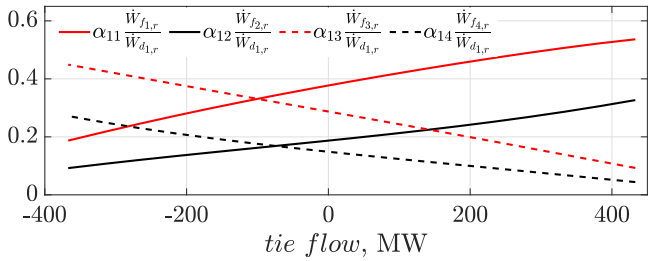}
  \caption{\textcolor{black}{Fractions in which the damping power of G$1$ is distributed as power dissipation in the windings of  G$1$ $-$ G$4$ for $0.69$ Hz mode.} }
  \label{distri_plot_4mc_1}
\end{figure}

\begin{figure}[h]
      \includegraphics[width=\linewidth]{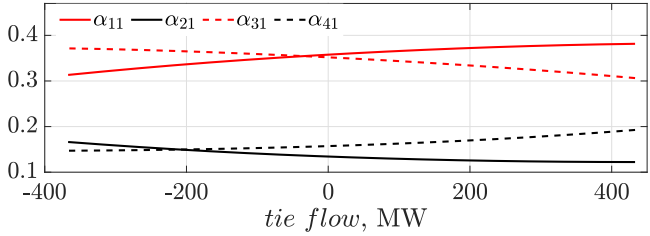}
  \caption{\textcolor{black}{Distribution of the power dissipation in G$1$ in the damping powers of G$1$ $-$ G$4$ for $0.69$ Hz mode.} }
  \label{distri_plot_4mc_2}
   \vspace{-0.2cm}
\end{figure}

 Next, in Fig. \ref{distri_plot_4mc_1}, the relative contributions from the power dissipations in the windings of G$1$ to G$4$ in constituting the damping power of G$1$, as discussed in Section \ref{distri}, are 
plotted for the $0.69$ Hz mode . Observe that, at a given operating point, the fractions $\alpha_{1j} \frac{\dot W_{f_{j,r}}}{\dot W_{d_{1,r}}}$ for $j = 1$ to $4$ add to $1$. In Fig. \ref{distri_plot_4mc_2} the distribution factors $\alpha_{i1}$-s for $i = 1$ to $4$ are shown. These describe fractions in which the power of oscillatory energy dissipation in G$1$ is distributed in the damping powers of G$1$ to G$4$. It can be seen that at any operating point, $\sum_{i=1}^{4}\alpha_{i1} = 1$. 
Since the machines are nearly identical and the network is symmetric with respect to the generators in this system, the nature of the distribution factors are similar for other generators, and thus are not repeated here. 
\par
\begin{figure}[]
\vspace{-0.3cm}
\begin{subfigure}{\linewidth}
\centering
\includegraphics[width = 1\linewidth]{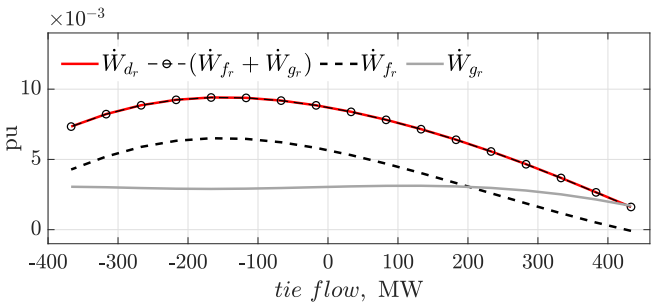}
\end{subfigure}
    \caption{Equality of total damping power with sum of average power dissipation in windings of all generators across different operating points in detailed $4-$machine system model
    for the $0.67$ Hz mode.}
    \label{consis_4mc_2}
\end{figure}

 \textit{(b) Detailed model:} 
The detailed model considers DC$1$A excitation system \cite{kundur} for each generator over and above the assumptions mentioned earlier -- the poorly-damped mode is shown in Table I. 
\textcolor{black}{For any mode, the power dissipation in the system is the summation of total power dissipations in field winding $\dot W_{f_{r}}$ and in damper winding $\dot W_{g_{r}}$ (see, (\ref{wfex_i}) and (\ref{wg_i}), respectively), where $\dot W_{f_{r}} = \sum_{i = 1}^{n_g}~ \dot W_{f_{i,r}}$ and $\dot W_{g_{r}} = \sum_{i = 1}^{n_g}~ \dot W_{g_{i,r}}$ 
}
For the detailed model, the consistency of total power dissipation with the total damping power is illustrated in Fig. \ref{consis_4mc_2}. 
\vspace{0.1cm}
\par
\begin{figure}[]
    \centering
     \vspace{-0.4cm}
    \includegraphics[width = 1.07\linewidth]{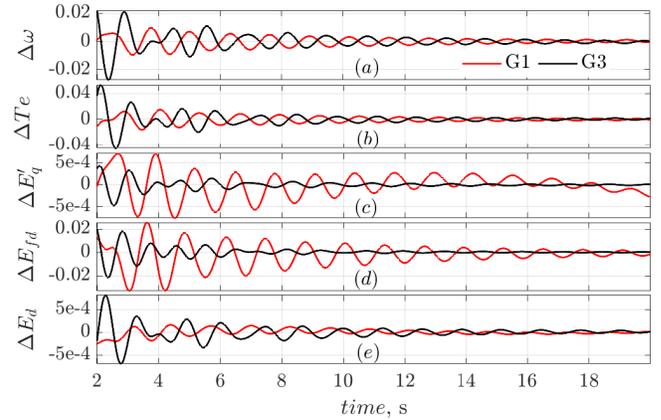}
    \vspace{-0.4cm}
    \caption{\textcolor{black}{Detrended post-fault time-domain plots of the state and output variables of G$1$ and G$3$.}}
    \label{dt_plot}
    \vspace{-0.3cm}
\end{figure}

 \textcolor{black}{\textit{$(c)$ Validation under large disturbances:}} \textcolor{black}{
Next, we consider the $4^{\text{th}}$-order machine model with IEEE ST$1$A excitation system for validation under large disturbances. Additionally, G$1$ is equipped with a PSS. 
}
     \textcolor{black}{We simulate a $5$-cycle three-phase self-clearing fault at $t = 1$ s near bus $8$. The detrended post-fault time-domain plots of the state and output variables of the two generators G$1$ and G$3$ are shown in Fig. \ref{dt_plot}. We obtain the relative modeshapes of these signals using the approach described in \cite{Dosiek}.}
\par \textcolor{black}{ We use $\Delta \omega_1$ as the reference signal and compute the modeshapes for all $\Delta \omega_i$, $\Delta T_{e_i}$, $\Delta E'_{q_i}$, $\Delta E'_{d_i}$, and $\Delta E_{fd_i}$-s for 
the critical mode in Table I. The damping and dissipative powers of all $4$ generators as computed using these modeshapes are shown in Table \ref{4mc_fault}.}
\textcolor{black}{It can be seen that, although for individual machines the values of $\dot W_{d_{i,r}}$-s are different from that of $\dot W_{f_{i,r}}$-s, their sum totals are almost equal. Also, these values nearly match those calculated from the small-signal model. }

\begin{table}[]
\centering
\vspace{0.3cm}
\caption{\textcolor{black}{DAMPING AND DISSIPATIVE POWERS IN DETAILED $4-$MACHINE SYSTEM MODEL FOR $0.77$ Hz MODE.}}
\label{4mc_fault}
\setlength{\tabcolsep}{8pt}
\renewcommand{\arraystretch}{1.3}
\begin{tabular}{c||c|c||c|c}
\hline 
\multicolumn{1}{c||}{\multirow{2}{*}{}} & \multicolumn{2}{c||}{\begin{tabular}[c]{@{}c@{}}Using the eigenvectors\\ obtained from the \\ small-signal model\end{tabular}} & \multicolumn{2}{c}{\begin{tabular}[c]{@{}c@{}}Using the modeshapes \\ estimated from the \\ time-domain responses\end{tabular}} \\ 
\multicolumn{1}{c||}{} & \makecell{ $\frac{ \dot W_{d_{i,r}}}{2  ~|\hat{c}_r|^2} ~$} & \makecell{ $\frac{\dot W_{f_{i,r}}}{2  ~|\hat{c}_r|^2} ~$} & \makecell{ $\frac{\dot W_{d_{i,r}}}{2  ~|\hat{c}_r|^2} ~$} & \makecell{$\frac{\dot W_{f_{i,r}}}{2  ~|\hat{c}_r|^2} ~$} \\ \hline \hline
G$1$ & $0.0382$ & $0.2294$ & $0.0361$ & $0.2274$ \\
G$2$ & $0.0321$ & $- 0.1830$ & $0.0340$ & $- 0.1877$ \\
G$3$ & $0.0435$ & $0.0794$ & $0.0405$ & $0.0754$ \\
G$4$ & $0.0521$ & $0.0401$ & $0.0472$ & $0.0390$ \\ \hline
Sum & $0.1659$ & $0.1659$ & $0.1578$ & $0.1541$ \\ \hline
\end{tabular}
\vspace{-0.4cm}
\end{table}

\begin{table}[H]
\centering
\setlength{\tabcolsep}{4pt}
\renewcommand{\arraystretch}{1.5}
\caption{\textcolor{black}{POORLY-DAMPED MODES IN 16-MACHINE SYSTEM}}
\label{16mc_eigen}
\begin{tabular}{c||c||c|c}
\hline
Machine Model & \begin{tabular}[c]{@{}c@{}}Eigenvalue\\ $\lambda_r  = \sigma_r + j\omega_{d_r}$\end{tabular} & \begin{tabular}[c]{@{}c@{}}Modal freq.\\ $f_r$ (Hz)\end{tabular} & \begin{tabular}[c]{@{}c@{}}Damping ratio\\ $\zeta_r$\end{tabular} \\ \hline \hline
\multirow{4}{*}{Simplified model} & $-0.0336 \pm j 2.1031$ & $0.34$ & $0.016$ \\
 & $-0.0338 \pm j 3.1288$ & $0.50$ & $0.011$ \\
 & $-0.1062 \pm j 3.7500$ & $0.60$ & $0.028$ \\
 & $-0.0264 \pm j 4.1031$ & $0.65$ & $0.006$ \\ \hline
\multirow{3}{*}{Detailed model} & $-0.0656  \pm j 3.1137$ & $0.49$ & $0.021$ \\
 & $-0.0981 \pm j 3.5169$ & $0.56$ & $0.028$ \\
 & $-0.1827 \pm j 4.9627$ & $0.79$ & $0.037$ \\ \hline
\end{tabular}
\end{table}

\subsection{IEEE $5-$area $16-$machine NY-NE Test System}
\begin{figure}[H]
    \centering
    \includegraphics[width=0.92\linewidth]{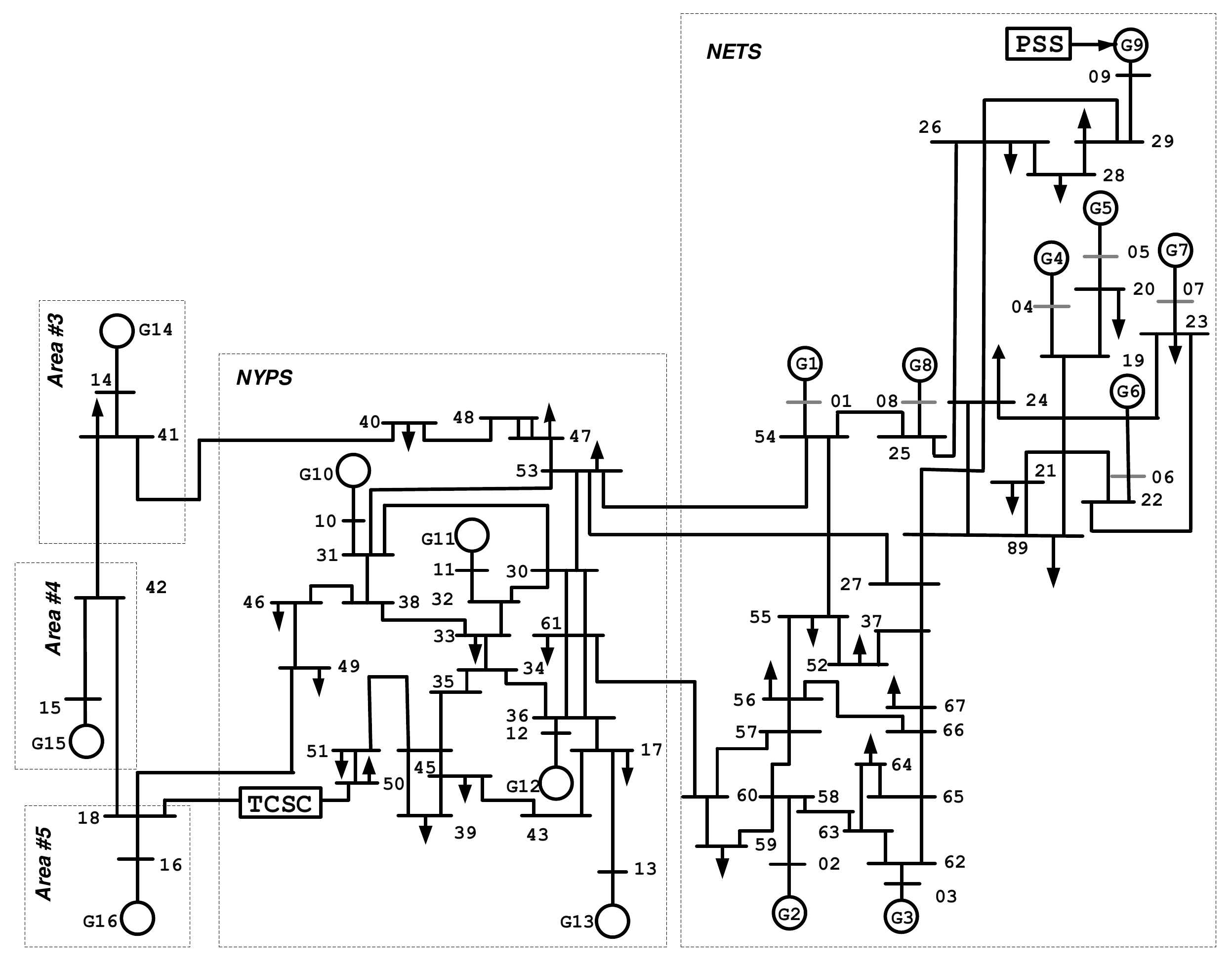}
    \caption{Single-line diagram of IEEE $5-$area $16-$machine NY-NE test system.}
    \vspace{-0.3cm}
    \label{fig_16mc}
\end{figure}

\begin{figure}[h]
\begin{subfigure}{\linewidth}
\centering
\vspace{-0.2cm}
\includegraphics[width = 1\linewidth]{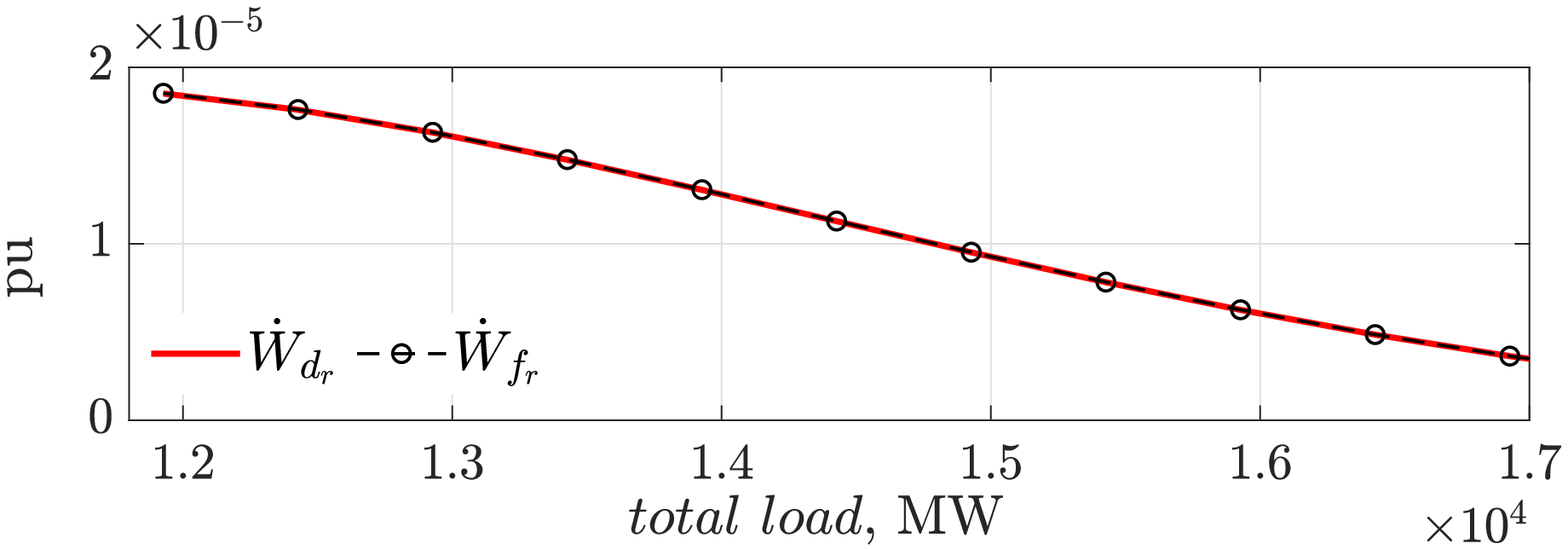}
\end{subfigure}
    \caption{
    Equality of total damping power with sum of average power dissipation in field windings of all generators across different operating points in simplified $16-$machine system model for $0.50$ Hz mode.}
    \label{16mc_consis_1}
    \vspace{-0.45cm}
\end{figure}

\begin{figure}[h]
\includegraphics[width = 1\linewidth]{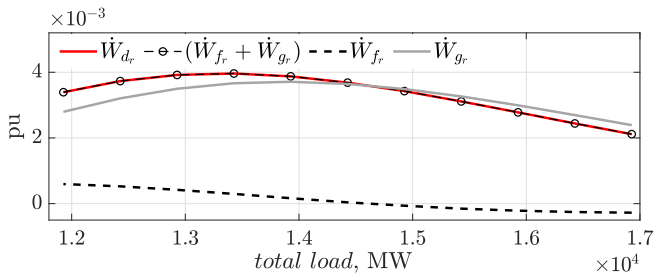}
    \caption{Equality of total damping power with sum of average power dissipation in windings of all generators across different operating points in detailed $16-$machine system model
    for $0.56$ Hz mode.}
    \label{16mc_consis_2}
\end{figure}

Next, consider the 
$16-$machine New York-New England test system shown in Fig. \ref{fig_16mc}. In the detailed model, G$1$ $-$ G$8$ have DC$1$A exciters, G$9$ is equipped with a ST$1$A exciter and a power system stabilizer (PSS), and the remaining generators have manual excitation. The machine and the network data can be obtained from \cite{nilthesis}. \textcolor{black}{The poorly damped modes of the system under nominal loading, for both simplified and detailed models, are shown in Table. \ref{16mc_eigen}.} Different operating points are obtained by uniformly changing the total system load. As before, in Figs \ref{16mc_consis_1} and \ref{16mc_consis_2}, the consistency of total damping power and sum of power dissipation in machine windings is shown respectively for the simplified and the detailed model -- each corresponding to a particular mode. \textcolor{black}{The negative values of $\dot W_{f_r}$ in Fig.\ref{16mc_consis_2} indicate that the excitation systems are contributing towards negative damping for higher loadings.}
 \vspace{-0.2cm}

\par
\begin{figure}[]
    \centering
    \vspace{-0.3cm}
    \includegraphics[width=1.04\linewidth]{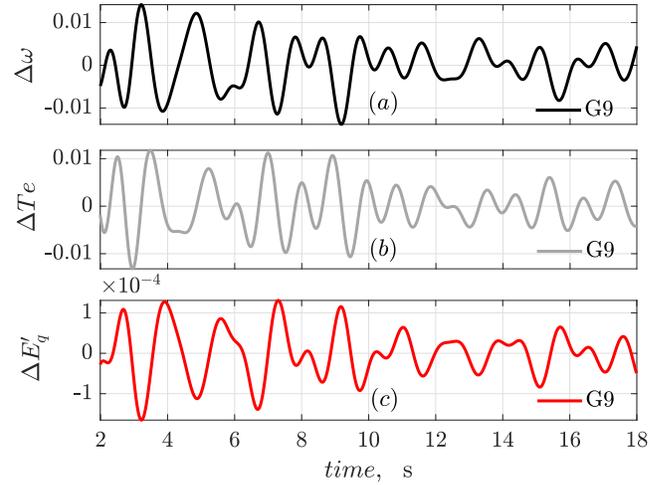}
    \vspace{-0.6cm}
    \caption{\textcolor{black}{
    Detrended time-domain responses of $(a)$ $\omega$, $(b)$ $T_e$, and $(c)$ $E'_q$ of G$9$ following a pulse disturbance in the excitation system of all generators in the $16-$machine system.}}
    \label{pulse_9}
    \vspace{-0.3cm}
\end{figure}

\begin{figure}[h]
    \centering
    \includegraphics[width = \linewidth]{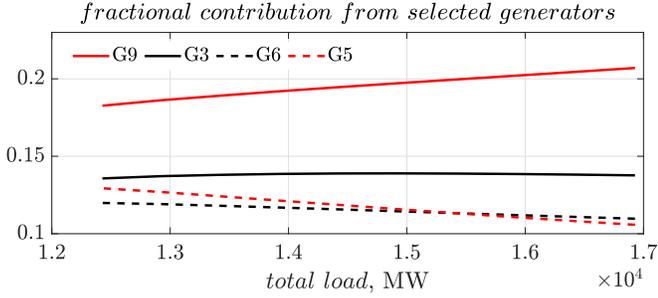}
    \caption{\textcolor{black}{Fractions in which the damping power of G$9$ is derived from the power dissipations in the windings of selected generators in NETS: G$3$, G$5$, G$6$, and G$9$ for the $0.60$ Hz mode.}}
    \label{frac_16_r3_variation}
    \vspace{-0.2cm}
\end{figure}

 \textcolor{black}{Next, we present the following case study for the simplified model to validate our claims regarding the distribution of damping power. In this study, for G$9$, we compare the values of $\frac{\alpha_{9j}~\dot W_{f_{j,r}}}{\dot W_{d_{9,r}}}$ (fractions in which the damping power of G$9$ is derived from the power dissipations in the windings of any $j^{th}$ generator) as calculated from small-signal model and as estimated from time-domain responses. }\textcolor{black}{Under nominal loading condition, a $0.2$ s pulse disturbance is applied to the excitation system of all generators and the time-domain responses of the state and output variables are obtained. Detrended plots of the variables of G$9$ are shown in Fig. \ref{pulse_9}. Using the approach in \cite{Dosiek} as before, we next estimate the relative modeshapes of $\omega_{_9}$, $T_{e_9}$, and $E'_{q_9}$, along with $E'_{q_i}$ for selected generators: G$3$, G$5$, and G$6$ from NETS, and G$11$ from NYPS from their time-domain responses.}
\textcolor{black}{We use $\Delta \omega_{_9}$ as the reference and the modeshapes are estimated for the $0.6$ Hz mode. \kc{For this mode, the generators of NETS oscillate against those in NYPS.} 
Next, using the estimated modeshapes, we compute the damping and dissipative powers of the selected generators as shown in Table \ref{damp_modeshape_16_r3}. Finally, using the $\alpha_{ij}$-s from the expression in (30) and the values estimated in Table \ref{damp_modeshape_16_r3}, we compute the fractional contributions from these selected generators towards the damping power of G$9$. These are shown in Table \ref{frac_16_r3}. As seen, the estimated values match those calculated from small-signal model. The variation in these fractions for change in system loading are shown in Fig. \ref{frac_16_r3_variation}.} 

\begin{table}[H]
\centering
\caption{\textcolor{black}{DAMPING AND DISSIPATIVE POWERS FOR THE $0.6$ Hz MODE ESTIMATED FROM TIME-DOMAIN RESPONSES}}
\label{damp_modeshape_16_r3}
\setlength{\tabcolsep}{7pt}
\renewcommand{\arraystretch}{2.4}
\begin{tabular}{c||c||c || c||c||c }
\hline
$\frac{\dot W_{d_{9,r}}}{2 |\hat{c}_r|^2} ~$ & $\frac{\dot W_{f_{3,r}}}{2 |\hat{c}_r|^2}$ & $\frac{\dot W_{f_{5,r}}}{2 |\hat{c}_r|^2}$ & $\frac{\dot W_{f_{6,r}}}{2 |\hat{c}_r|^2}$ & $\frac{\dot W_{f_{9,r}}}{2 |\hat{c}_r|^2}$ & $\frac{\dot W_{f_{11,r}}}{2 |\hat{c}_r|^2}$  \\ \hline \hline
$0.0806$ & $0.0905$ & $0.0663$ & $0.0791$ & $0.1340$ & $0.0334$ \\ \hline 
\end{tabular}
\end{table}

\begin{table}[H]
\centering
\caption{\textcolor{black}{FRACTIONAL CONTRIBUTION FROM SELECTED GENERATORS TOWARDS DAMPING POWER OF G$9$ FOR THE $0.6$ Hz MODE}}
\label{frac_16_r3}
\setlength{\tabcolsep}{3.5pt}
\renewcommand{\arraystretch}{1.5}
\begin{tabular}{c||c|c|c|c|c}
\hline
frac. contr. from  & G$9$ & G$3$ & G$5$ & G$6$ & G$11$\\ \hline \hline
small-signal model & $0.2064$ & $0.1382$ & $0.1063$ & $0.1102$ & $0.0520$ \\ \hline
time-domain estimation & $0.2046$ & $0.1369$ & $0.1003$ & $0.1198$ & $0.0501$ \\ \hline 
\end{tabular}
\end{table}

\par
\begin{figure}[H]
      \includegraphics[width=\linewidth]{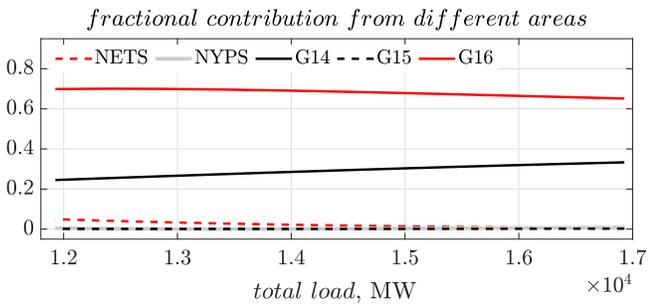}
  \caption{\textcolor{black}{Fractions in which the damping power of G$16$ is derived from the power dissipations in the windings of  generators in NETS (G$1$ $-$ G$9$), NYPS (G$10$ $-$ G$13$), and G$14$ $-$ G$16$ for the $0.50$ Hz mode.} }
  \label{distri_plot_16mc}
  \vspace{-0.1cm}
\end{figure}

 \kc{Now, we consider the other poory-damped mode at $0.5$ Hz, for which generators outside NETS $-$ G$14$ and G$16$ oscillate against each other with marginal participation from generators in other areas. The damping and dissipative powers of these two generators, for the mode, are shown in Table \ref{14_16_powers}. Note that, since other generators do not participate in the mode, the summation of damping powers of G$14$ and G$16$ is approximately equal to the summation of their dissipative powers.} Finally, in Fig. \ref{distri_plot_16mc}, the relative dissipative contributions from the generators in different areas in constituting the damping power of G$16$ are shown. For G$1$ $-$ G$9$ and G$10$ $-$ G$13$, their aggregates
\begin{equation*}
    ~~\sum_{j=1}^9 \alpha_{16,j} \frac{\dot W_{f_{j,r}}}{\dot W_{d_{16,r}}} ~~~\text{and}~ ~~\sum_{j=10}^{13} \alpha_{16,j} \frac{\dot W_{f_{j,r}}}{\dot W_{d_{16,r}}}
    \vspace{-0.05cm}
    \end{equation*}
     are shown as contributions from NETS and NYPS, respectively. \kcc{We note that G$14$ and G$16$ have the highest participation in the mode, which is aligned with the observation that these generators have the highest dissipative contribution.}

\begin{table}[H]
\centering
\caption{\textcolor{black}{DAMPING AND DISSIPATIVE POWERS OF G$14$ and G$16$ FOR THE $0.5$ Hz MODE}}
\label{14_16_powers}
\setlength{\tabcolsep}{8pt}
\renewcommand{\arraystretch}{1.3}
\begin{tabular}{c||c|c||c|c}
\hline 
\multicolumn{1}{c||}{\multirow{2}{*}{}} & \multicolumn{2}{c||}{\begin{tabular}[c]{@{}c@{}} Calculated from \\ small-signal model\end{tabular}} & \multicolumn{2}{c}{\begin{tabular}[c]{@{}c@{}} Estimated from the \\ time-domain responses\end{tabular}} \\ 
\multicolumn{1}{c||}{} & \makecell{ $\frac{ \dot W_{d_{i,r}}}{2  ~|\hat{c}_r|^2} ~$} & \makecell{ $\frac{\dot W_{f_{i,r}}}{2  ~|\hat{c}_r|^2} ~$} & \makecell{ $\frac{\dot W_{d_{i,r}}}{2  ~|\hat{c}_r|^2} ~$} & \makecell{$\frac{\dot W_{f_{i,r}}}{2  ~|\hat{c}_r|^2} ~$}  \\ \hline \hline
G$14$ & $0.2150$ & $0.1642$ & $0.2070$ & $0.1503$ \\
G$16$ & $0.2665$ & $0.3212$ & $0.2904$ & $0.3418$ \\ \hline
Sum & $0.4815$ & $0.4854$ & $0.4974$ & $0.4921$ \\ \hline
\end{tabular}
\end{table}

\section{Conclusions}
A mathematical proof for the equality of total damping power of the system and the total power dissipation in generators was presented for multimachine systems. It was demonstrated that the equality while true when added over all generators, does not hold for individual machines. Thereafter, distribution factors were derived representing the dissipative contributions from different generators in constituting the damping power of each machine.
\vspace{-0.3cm}
\begin{appendices} \renewcommand{\thesectiondis}[2]{\Alph{section}:}
  \section{Phasor Notation}
  \label{phasornot}
  Let the small-signal dynamics of an autonomous system be represented by the state space model $\dot{\mathbf{x}}(t) = \mathbf{A} \mathbf{ x}(t)$. Next, assuming there are $m$ oscillatory modes in the response, each due to a complex-conjugate eigenvalue pair $\lambda_r$ ($= \sigma_r + j\omega_{d_r}$) and $\lambda_r^*$ of $\mathbf{A}$, the time evolution of any $i^{\text{th}}$ state variable $x_i(t)$ can be expressed as the sum of $m$ modal constituents, as shown in eqn (\ref{resp}) 
\begin{equation}
\small
\begin{aligned}
    \label{resp}
    x_i(t) = \sum_{r=1}^{m}x_{i,r}(t) &= \sum_{r=1}^{m} \Big\{e^{\lambda_rt}c_r\psi_{i,r} + e^{\lambda_r^*t}c_r^{*}\psi^{*}_{i,r} \Big\}
    \end{aligned}
\end{equation}
where, $c_r = \boldsymbol{\phi}_r^T\mathbf{x}(0)$, and $\boldsymbol{\phi}_r^T$ and $\Psi_r$ are respectively the left and right eigenvectors of $\mathbf{A}$ corresponding to the eigenvalue $\lambda_r$ with $\psi_{i,r}$ as the $i^{\text{th}}$ entry of $\Psi_r$.
\noindent \par Denoting $2 c_r \psi_{i,r} \overset{\Delta}{=} \beta_{i,r} e^{j\gamma_{i,r}}$, $x_{i,r}(t)$ reduces to
\begin{equation}
\small
\label{beta}
   x_{i,r}(t) = 2~\Re \big \{e^{\lambda_rt}c_r\psi_{i,r} \big\} = \beta_{i,r}~e^{\sigma_r t}\cos(\omega_{d_r}t + \gamma_{i,r}) .
\end{equation}
\par This sinusoidal variation is represented in the 
\kc{dynamic} phasor \kc{(mentioned as `phasor' going forward)} notation using the magnitude and phase of the signal, as shown in eqn (\ref{phasdef}). 
\begin{equation}
\small
\label{phasdef}
 \textcolor{black}{    \vec{x}_{i,r}(t) ~\overset{\Delta}{=}~ \beta_{i,r}~ e^{\sigma_rt}~\angle{\gamma_{i,r}} ~=~ 2~c_r~ \psi_{i,r}~e^{\sigma_r t}}
\end{equation}
\textcolor{black}{$\vec{x}_{i,r} (t)$ is a phasor rotating at the modal frequency $\omega_{d_r}$ with its amplitude having an exponential decay.
It represents the time evolution of $x_{i,r}(t)$.} We denote $\vec{\mathbf{x}}_r$ as the vector of $\vec{x}_{i,r} (t)$-s.
\par \textcolor{black}{Next, from (\ref{beta}),}
\begin{equation}
\begin{aligned}
\small
\label{beta_dot}
 \textcolor{black}{ \dot  x_{i,r}(t) =  \beta_{i,r}~e^{\sigma_r t}~\Big\{ \sigma_r~\cos( } & \textcolor{black}{\omega_{d_r}t + \gamma_{i,r}) ~-}\\ & \textcolor{black}{ ~\omega_{d_r}~\sin(\omega_{d_r}t + \gamma_{i,r}) \Big\} } .
\end{aligned}
\end{equation}
\textcolor{black}{Considering the mode to be poorly-damped, $|\sigma_r|~<< \omega_{d_r}$, this reduces to $\dot  x_{i,r}(t) \approx -\beta_{i,r}~e^{\sigma_r t}~~\omega_{d_r}~\sin(\omega_{d_r}t + \gamma_{i,r})$. Therefore, }
\begin{equation}
\small
\label{phasdef_2}
 \textcolor{black}{  \vec { \dot{x}}_{i,r} (t) \approx j~\omega_{d_r}~\vec{x}_{i,r}(t)} .
\end{equation}

\section{Proof of Claims}
\label{poc}
\noindent Observe that, from (\ref{sys_int}), $\mathbf{M}$, $\mathbf{N}$, $\mathbf{C}$, and $\mathbf{D}$ can be structured as 
\begin{equation}
\begin{aligned}
\small
\mathbf{M} = \left[ \begin{array}{c c c} \mathbf{M_{11}}  & \mathbf{M_{12}}  & \mathbf{M_{13}} \\  \mathbf{M_{21}} & \mathbf{M_{22}}  & \mathbf{M_{23}} \\ \mathbf{M_{31}} & \mathbf{M_{32}}  & \mathbf{M_{33}} \end{array}\right] ~~~ \mathbf{N} = \left[ \begin{array}{c c } \mathbf{N_{11}}  & \mathbf{N_{12}}   \\  \mathbf{N_{21}} & \mathbf{N_{22}}  \\ \mathbf{N_{31}} & \mathbf{N_{32}}   \end{array}\right]\\  
\small \mathbf{C} = \left[ \begin{array}{c c c} \mathbf{C_{11}}  & \mathbf{C_{12}}  & \mathbf{C_{13}} \\  \mathbf{C_{21}} & \mathbf{C_{22}}  & \mathbf{C_{23}} \end{array}\right] ~~~ \small \mathbf{D} =  \arraycolsep=1.4pt\def\arraystretch{1.4}\left[ \begin{array}{c c} \mathbf{D}_{\mathbf{11}} & \mathbf{D}_{\mathbf{12}}\\  \mathbf{D}_{\mathbf{21}} & \mathbf{D}_{\mathbf{22}} \end{array}\right].
\end{aligned}
\end{equation}
\par Following the notation that, $\mathbf{D_{kl}}(i,j)$ is the $(i,j)^{\text{th}}$ element of the $(k,l)^{\text{th}}$ submatrix $\mathbf{D_{kl}}$ of $\mathbf{D}$, we may write $\forall~j \neq i$
\begin{subequations}
\small
\begin{align}
    \small
    \label{D12a} \mathbf{D_{12}}(i,j) &= V_{j_0}\frac{\partial f_i}{\partial V_j}\biggm|_0 = -V_{i_0}V_{j_0}Y_{ij}\sin(\theta_{i_0} - \theta_{j_0}) \\
    \label{D12b} \mathbf{D_{21}}(i,j) &= \frac{\partial g_i}{\partial \theta_j}\biggm|_0 =  V_{i_0}V_{j_0}Y_{ij}\sin(\theta_{i_0} - \theta_{j_0}) = \mathbf{D_{12}}(j,i)\\
    \label{D11}\mathbf{D_{11}}(i,j) &= \frac{\partial f_i}{\partial     \theta_j}\biggm|_0 =  V_{i_0}V_{j_0}Y_{ij}\cos(\theta_{i_0} - \theta_{j_0}) = \mathbf{D_{11}}(j,i)\\
    \begin{split} \small \label{D22}\mathbf{D_{22}}(i,j) &= V_{j_0}\frac{\partial g_i}{\partial V_j}\biggm|_0 = V_{i_0}V_{j_0}Y_{ij}\cos(\theta_{i_0} - \theta_{j_0})= \mathbf{D_{22}}(j,i).
    \end{split}
\end{align}
\end{subequations}
Similarly, it can be shown that,
\begin{equation}
    \small
    \label{D12d} \mathbf{D_{12}}(i,i) = V_{i_0}\frac{\partial f_i}{\partial V_i}\biggm|_0 =
     \frac{\partial g_i}{\partial \theta_i}\biggm|_0 
     = \mathbf{D_{21}}(i,i) .
\end{equation}
\par Therefore, from eqns (\ref{D12a}),  (\ref{D12b}), and (\ref{D12d}), it can be inferred that $\mathbf{D}^{^T}_{\mathbf{12}} = \mathbf{D_{21}}$. Additionally, from eqns (\ref{D11}) $-$ (\ref{D22}), $\mathbf{D}^{^T}_{\mathbf{11}} = \mathbf{D_{11}}$ and $\mathbf{D}^{^T}_{\mathbf{22}} = \mathbf{D_{22}}$. Thus, 
\begin{equation}
\small
    \mathbf{D}^{^T} =  \arraycolsep=1.4pt\def\arraystretch{1.4}\left[ \begin{array}{c c} \mathbf{D}_{\mathbf{11}} & \mathbf{D}_{\mathbf{12}}\\  \mathbf{D}_{\mathbf{21}} & \mathbf{D}_{\mathbf{22}} \end{array}\right]^T = \left[ \begin{array}{c c} \mathbf{D}^{^T}_{\mathbf{11}} & \mathbf{D}^{^T}_{\mathbf{21}}\\  \mathbf{D}^{^T}_{\mathbf{12}} & \mathbf{D}^{^T}_{\mathbf{22}} \end{array}\right]  = \mathbf{D} .
\end{equation}
Further, $\mathbf{D}$ being real and symmetric implies  $\mathbf{D}^{^{-1}}$ is also real and symmetric 
\begin{equation}
    \small
    \label{Dinv}
    \implies \mathbf{D}^{^{-T}} = \mathbf{D}^{^{-1}} .
\end{equation}
\par \textit{Proof of Claim (1)} : Recall, $\mathbf{A} = \mathbf{M - ND^{^{-1}}C}$. Therefore, 
\begin{equation}
\small
\label{prop1}
\begin{aligned}
  \mathbf{A_{33}} &= \mathbf{M_{33}} -  \left[ \begin{array}{c c} \mathbf{N}_{\mathbf{31}} & \mathbf{N}_{\mathbf{32}} \end{array}\right] \mathbf{D}^{^{-1}}  \left[ \begin{array}{c} \mathbf{C}_{\mathbf{13}} \\ \mathbf{C}_{\mathbf{23}} \end{array}\right]\\
  \implies \mathbf{A}^{^T}_{\mathbf{33}} &= \mathbf{M}^{^T}_{\mathbf{33}} -  \arraycolsep=1.4pt\def\arraystretch{1.4}\left[ \begin{array}{c c} \mathbf{C}^{^T}_{\mathbf{13}} & \mathbf{C}^{^T}_{\mathbf{23}} \end{array}\right] \mathbf{D}^{^{-T}}  \left[ \begin{array}{c} \mathbf{N}^{^T}_{\mathbf{31}} \\ \mathbf{N}^{^T}_{\mathbf{32}} \end{array}\right]
\end{aligned}
\end{equation}
From eqns (\ref{f3}) and (\ref{g1}) observe that, $\forall~j \neq i$,
\begin{equation}
\small
\label{c130}
\begin{aligned}
 \mathbf{N_{31}}(i,j) =  \frac{\partial \dot E_{q_i}'}{\partial \theta_j}\biggm|_0 = 0~;  ~~~~~~
  \mathbf{C_{13}}(i,j) =  \frac{\partial f_i}{\partial E_{q_j}'}\biggm|_0 = 0~;
 \end{aligned}
 \end{equation}
 and for elements on the principal diagonal,
\begin{subequations}
\small
\label{c13}
\begin{align}
\mathbf{N_{31}}(i,i) &=  \frac{\partial \dot E_{q_i}'}{\partial \theta_i}\biggm|_0 =  \frac{ V_{i_0}\sin(\delta_{i_0}-\theta_{i_0})}{x_{d_i}'} \bigg( \frac{x_{d_i}-x_{d_i}'}{T_{do_i}'} \bigg)\\
\mathbf{C_{13}}(i,i) &=  \frac{\partial f_i}{\partial E_{q_i}}\biggm|_0 =  \frac{ V_{i_0}\sin(\delta_{i_0}-\theta_{i_0})}{x_{d_i}'}
\end{align}
\end{subequations}
Further note, $\mathbf{N_{31}}$ and $\mathbf{C_{13}}$ are rectangular matrices of dimensions $ \mathbb{R}^{n_g\times n}$ and $ \mathbb{R}^{n\times n_g}$ respectively. Therefore, combining eqns (\ref{c130}) and (\ref{c13}) we get
\begin{equation}
\small
\label{c23a}
    \mathbf{P}^{^{-1}}\mathbf{C_{13}}^{^T} = \mathbf{N_{31}} 
\end{equation}
where, $\mathbf{P}$ is a diagonal matrix with $\mathbf{P}(i,i) = \frac{T_{do_i}'}{x_{d_i}-x_{d_i}'}$. 
\par \noindent Similarly, from eqns (\ref{f3}) and (\ref{g2}), $\forall~j \neq i$,
 \begin{equation}
 \notag
 \small
\begin{aligned}
\mathbf{N_{32}}(i,j) &=  V_{j_0}\frac{\partial \dot E_{q_i}'}{\partial V_j}\biggm|_0 = 0~;~~~
 \mathbf{C_{23}}(i,j) =  \frac{\partial g_i}{\partial E_{q_j}'}\biggm|_0 = 0~;~~ \text{and}
\end{aligned}
\end{equation}
\begin{equation*}
\small
\label{c23}
\begin{aligned}
\mathbf{N_{32}}(i,i) &=  V_{i_0}\frac{\partial \dot E_{q_i}'}{\partial V_j}\biggm|_0 =  \frac{ V_{i_0}\cos(\delta_{i_0}-\theta_{i_0})}{x_{d_i}'} \bigg( \frac{x_{d_i}-x_{d_i}'}{T_{do_i}'} \bigg)\\
\mathbf{C_{23}}(i,i) &=  \frac{\partial g_i}{\partial E_{q_i}}\biggm|_0 =  \frac{ V_{i_0}\cos(\delta_{i_0}-\theta_{i_0})}{x_{d_i}'}
\end{aligned}
\end{equation*}
Therefore, following arguments as before,
\begin{equation}
\label{c23b}
\small
   \mathbf{P}^{^{-1}}\mathbf{C_{23}}^{^T} = \mathbf{N_{32}} 
\end{equation}
\par \noindent Finally, observe that $\mathbf{M_{33}} \in \mathbb{R}^{n_g\times n_g}$ with $\mathbf{M_{33}}(i,j) =  \frac{\partial \dot E_{q_i}'}{\partial \delta_j}\biggm|_0 = 0~~ \forall~ j \neq i$. This implies $\mathbf{M_{33}}$ is diagonal.
\par Using the results (\ref{Dinv}), (\ref{c23a}) and (\ref{c23b}), eqn  (\ref{prop1}) can be rewritten as
\begin{equation}
\small
\notag
 \begin{aligned}
 \mathbf{P}^{^{-1}}\mathbf{A}^{^T}_{\mathbf{33}}\mathbf{P} &= \mathbf{P}^{^{-1}}\mathbf{M}^{^T}_{\mathbf{33}}\mathbf{P} - \mathbf{P}^{^{-1}} \arraycolsep=1.4pt\def\arraystretch{1.4}\left[ \begin{array}{c c} \mathbf{C}^{^T}_{\mathbf{13}} & \mathbf{C}^{^T}_{\mathbf{23}} \end{array}\right] \mathbf{D}^{^{-T}}  \left[ \begin{array}{c} \mathbf{N}^{^T}_{\mathbf{31}} \\ \mathbf{N}^{^T}_{\mathbf{32}} \end{array}\right]\mathbf{P}\\
 &= \mathbf{M_{33}} -  \left[ \begin{array}{c c} \mathbf{N}_{\mathbf{31}} & \mathbf{N}_{\mathbf{32}} \end{array}\right] \mathbf{D}^{^{-1}}  \left[ \begin{array}{c} \mathbf{C}_{\mathbf{13}} \\ \mathbf{C}_{\mathbf{23}} \end{array}\right] = \mathbf{A}_{\mathbf{33}} .
 \end{aligned}   
\end{equation}
This concludes the proof. \qedsymbol\\
\par \textit{Proof of Claim (2)} : It can be seen from eqns (\ref{f2}) and (\ref{f3}) that blocks $\mathbf{M_{31}}$ and $\mathbf{M_{23}}$ are diagonal. Also, 
\begin{subequations}
\small
\begin{align}
    \mathbf{M_{31}}(i,i) &= \frac{\partial \dot E_{q_i}'}{\partial \delta_{i}}\biggm|_0 = -\frac{V_{i_0}\sin(\delta_i-\theta_i)}{x_{d_i}'}\bigg(\frac{x_{d_i} - x_{d_i}'}{T_{do_i}'}\bigg)\\
    \mathbf{M_{23}}(i,i) &= \frac{\partial \dot \omega_i}{\partial E_{q_i}'}\biggm|_0 = -\frac{V_{i_0}\sin(\delta_i-\theta_i)}{2 H_i~x_{d_i}'}~\omega_s
\end{align}
\end{subequations}
Therefore, we my write 
\begin{equation}
\small
    \label{m23}
    \mathbf{M}^{^T}_{\mathbf{31}}~ \mathbf{P} = \frac{2~ \mathbf{H}}{\omega_s} ~\mathbf{M_{23}}
\end{equation}
where $\mathbf{H}$ is diagonal with $\mathbf{H}(i,i) = H_i$.
\par \noindent Now, as before, for $\mathbf{N}$ and $\mathbf{C}$ matrices, 
\begin{equation*}
\small
\label{c110}
\begin{aligned}
\mathbf{N_{21}}(i,j) =  \frac{\partial \dot \omega_i}{\partial \theta_j}\biggm|_0 = 0~;  ~~~~~
 \mathbf{C_{11}}(i,j) =  \frac{\partial f_i}{\partial \delta_j}\biggm|_0 = 0. ~~~\text{Also,}
 \end{aligned}
 \end{equation*}
 \begin{equation*}
 \small
 \begin{aligned}
  \mathbf{N_{21}}(i,i) =  \frac{\partial \dot \omega_i}{\partial \theta_i}\biggm|_0 &= \frac{\omega_s~E_{q_{i_0}}'V_{i_0}\cos(\delta_{i_0} - \theta_{i_0})}{2H_i~x_{d_i}'} \\&- \frac{\omega_s V_{i_0}^2\sin2(\delta_{i_0}-\theta_{i_0})}{2H_i}\Big(\frac{x_{q_i} - x_{d_i}'}{x_{q_i}x_{d_i}'}\Big)\\
  \mathbf{C_{11}}(i,i) =  \frac{\partial f_i}{\partial \delta_i}\biggm|_0 &= \frac{E_{q_{i_0}}'V_{i_0}\cos(\delta_{i_0} - \theta_{i_0})}{x_{d_i}'} \\- V_{i_0}^2\sin2(&\delta_{i_0}-\theta_{i_0})\Big(\frac{x_{q_i} - x_{d_i}'}{x_{q_i}x_{d_i}'}\Big) = \frac{2~H_i}{\omega_s}~\mathbf{N_{21}}(i,i)
 \end{aligned}
 \end{equation*}
 Combining these with the fact that, $\mathbf{N_{21}} \in \mathbb{R}^{n_g\times n}$ and $\mathbf{C_{11}} \in \mathbb{R}^{n\times n_g}$ we may write, 
 \begin{equation}
 \label{c11a}
 \small
     \mathbf{C}^{^T}_{\mathbf{11}} = \frac{2~\mathbf{H}}{\omega_s}~\mathbf{N_{21}}. 
 \end{equation}
 Similarly, it can be shown that
 \begin{equation}
 \small
 \label{c11b}
     \mathbf{C}^{^T}_{\mathbf{21}} = \frac{2~\mathbf{H}}{\omega_s}~\mathbf{N_{22}}.
 \end{equation}
 Now, recall $\mathbf{A} = \mathbf{M - ND^{^{-1}}C}$. Therefore, 
\begin{equation}
\small
\label{prop2}
\begin{aligned}
  \mathbf{A_{31}} &= \mathbf{M_{31}} -  \left[ \begin{array}{c c} \mathbf{N}_{\mathbf{31}} & \mathbf{N}_{\mathbf{32}} \end{array}\right] \mathbf{D}^{^{-1}}  \left[ \begin{array}{c} \mathbf{C}_{\mathbf{11}} \\ \mathbf{C}_{\mathbf{21}} \end{array}\right]\\
  \implies \mathbf{A}^{^T}_{\mathbf{31}}\mathbf{P} &= \mathbf{M}^{^T}_{\mathbf{33}}\mathbf{P} -  \arraycolsep=1.4pt\def\arraystretch{1.4}\left[ \begin{array}{c c} \mathbf{C}^{^T}_{\mathbf{11}} & \mathbf{C}^{^T}_{\mathbf{21}} \end{array}\right] \mathbf{D}^{^{-T}}  \left[ \begin{array}{c} \mathbf{N}^{^T}_{\mathbf{31}} \\ \mathbf{N}^{^T}_{\mathbf{32}}  \end{array}\right]\mathbf{P}
\end{aligned}
\end{equation}
Next, substituting eqns (\ref{Dinv}) and (\ref{c23a}) $-$ (\ref{c11b}) in (\ref{prop2})
\begin{equation}
\small
\label{prop2b}
\begin{aligned}
 \mathbf{A}^{^T}_{\mathbf{31}}\mathbf{P} &= \frac{2~\mathbf{H}}{\omega_s}~\mathbf{M_{23}}~ - ~ \arraycolsep=1.4pt\def\arraystretch{1.2}\frac{2~\mathbf{H}}{\omega_s}~\left[ \begin{array}{c c} \mathbf{N}_{\mathbf{21}} & \mathbf{N}_{\mathbf{22}} \end{array}\right] \mathbf{D}^{^{-1}}  \left[ \begin{array}{c} \mathbf{C}_{\mathbf{13}} \\ \mathbf{C}_{\mathbf{23}}  \end{array}\right]\\
  &= \frac{2~\mathbf{H}}{\omega_s}~\mathbf{A_{23}}
\end{aligned}
\end{equation}
This concludes the proof. \qedsymbol \vspace{0.2cm}
\par \textit{Proof of Claim (3)} : As before, observe from eqn (\ref{f2}) that $\mathbf{M_{21}}$ is diagonal. Therefore, we may write
\begin{equation}
    \small
    \label{m21}
    \mathbf{H~M_{21}} = \mathbf{M}^T_{\mathbf{21}}~\mathbf{H} .
\end{equation}
\begin{equation}
\small
\label{prop3a}
\begin{aligned}
\text{Also,~~~~~~}~\small \mathbf{A_{21}} &= \mathbf{M_{21}}~ - ~ \left[ \begin{array}{c c} \mathbf{N}_{\mathbf{21}} & \mathbf{N}_{\mathbf{22}} \end{array}\right] \mathbf{D}^{^{-1}} \arraycolsep=1.4pt\def\arraystretch{1.0}
\left[ \begin{array}{c} \mathbf{C}_{\mathbf{11}} \\ \mathbf{C}_{\mathbf{21}} \end{array}\right]\\
  \implies 2~\mathbf{A}^{^T}_{\mathbf{21}}\mathbf{H} &= \mathbf{M}^{^T}_{\mathbf{21}}\mathbf{H} ~-~  \arraycolsep=1.4pt\def\arraystretch{1.4}\left[ \begin{array}{c c} \mathbf{C}^{^T}_{\mathbf{11}} & \mathbf{C}^{^T}_{\mathbf{21}} \end{array}\right] \mathbf{D}^{^{-T}}  \left[ \begin{array}{c} 2~\mathbf{N}^{^T}_{\mathbf{21}}~\mathbf{H}\\ 2~\mathbf{N}^{^T}_{\mathbf{22}}~\mathbf{H}  \end{array}\right]
\end{aligned}
\end{equation}
Finally, substituting eqns (\ref{Dinv}), (\ref{c11a}), (\ref{c11b}), and (\ref{m21}) in (\ref{prop3a}) 
\begin{equation}\small
\begin{aligned}
    \small
   \small  2~\mathbf{A}^{^T}_{\mathbf{21}}\mathbf{H} &= 2~\mathbf{H}~\mathbf{M_{21}}~ - ~ 2~\mathbf{H}~\arraycolsep=1.4pt\def\arraystretch{1.0}\left[ \begin{array}{c c} \mathbf{N}_{\mathbf{21}} & \mathbf{N}_{\mathbf{22}} \end{array}\right] \mathbf{D}^{^{-1}} \arraycolsep=1.4pt\def\arraystretch{1.0}\left[ \begin{array}{c} \mathbf{C}_{\mathbf{11}} \\ \mathbf{C}_{\mathbf{21}} \end{array}\right] \\\small &= 2~\mathbf{H}~\mathbf{A_{21}} .
    \end{aligned}
\end{equation}
This concludes the proof. \qedsymbol \\
\par \textit{Proof of Claim (4)} : From 
the definition of $\mathbf{K}_r$ and (\ref{Temm}),
\begin{equation*}
    \small
    \label{Ksym1}
    \begin{aligned}
      \mathbf{K}^T_r = -\bigg \{\frac{2}{\omega_s}~\mathbf{A}^{^T}_{\mathbf{21}}\mathbf{H} + \mathbf{A}^{^T}_{\mathbf{31}}~(j\omega_{d_r}\mathbf{I}-\mathbf{A}^{^T}_{\mathbf{33}})^{^{-1}}(\frac{2}{\omega_s}~\mathbf{A}^{^T}_{\mathbf{23}}~\mathbf{H})\bigg\}\frac{1}{j\omega_{d_r}}\\
      \begin{split} =  -\bigg \{ \frac{2}{\omega_s}~\mathbf{A}^{^T}_{\mathbf{21}}\mathbf{H} + \mathbf{A}^{^T}_{\mathbf{31}}~\mathbf{P}~(j\omega_{d_r}\mathbf{I}~-~ &\mathbf{P}^{-1}\mathbf{A}^{^T}_{\mathbf{33}}\mathbf{P})^{^{-1}}\\&\mathbf{P}^{-1}(\frac{2~\mathbf{H}}{\omega_s}~\mathbf{A_{23}})^T\bigg\}\frac{1}{j\omega_{d_r}}\end{split}
    \end{aligned}
\end{equation*}
Next, using claims (1) $-$ (3) we can re-write $\mathbf{K}^T_r$ as
\begin{equation}
    \small
    \label{Ksym2}
    \begin{aligned}
     \mathbf{K}^T_r = -\frac{2~\mathbf{H}}{\omega_s}~\bigg \{ \mathbf{A_{21}} + \mathbf{A_{23}}~(j\omega_{d_r}\mathbf{I}-\mathbf{A_{33}})^{^{-1}}\mathbf{A_{31}}\bigg\}~\frac{1}{j\omega_{d_r}} = \mathbf{K}_r
    \end{aligned}
\end{equation}
Now, $\forall~\mathbf{x}~\in~\mathbb{C}^{n_g}$, let us decompose it into its real and imaginary parts as shown: $\mathbf{x} = \mathbf{x}_1 + j\mathbf{x}_2$. 
Therefore,
\begin{equation}
    \small
    \label{cl41}
    \begin{aligned}
       \mathbf{x}^H \mathbf{K}_r \mathbf{x}  = (\mathbf{x}_1^T - j\mathbf{x}_2^T) \Big(\Re (\mathbf{K}_r) + j\Im(\mathbf{K}_r) \Big) (\mathbf{x}_1 + j\mathbf{x}_2) 
    \end{aligned}
\end{equation}
Further, using eqn (\ref{Ksym2}), we can infer on the symmetry of both the real and imaginary parts of $\mathbf{K}_r$. Hence,
\begin{equation}
\label{symmetry}
\small
    \begin{aligned}
      \mathbf{x}_1^T \Im(\mathbf{K}_r)\mathbf{x}_2 = \mathbf{x}_2^T \Im(\mathbf{K}_r)\mathbf{x}_1 ~ \text{and} ~
      \mathbf{x}_1^T \Re(\mathbf{K}_r)\mathbf{x}_2 = \mathbf{x}_2^T \Re(\mathbf{K}_r)\mathbf{x}_1 .
    \end{aligned}
\end{equation}
This reduces the real part of eqn (\ref{cl41}) as follows
\begin{equation}
    \small
    \begin{aligned}
      \Re \{ \mathbf{x}^H \mathbf{K}_r \mathbf{x} \} = \mathbf{x}_1^T \Re(\mathbf{K}_r)\mathbf{x}_1 + \mathbf{x}_2^T \Re(\mathbf{K}_r)\mathbf{x}_2 .
    \end{aligned}
\end{equation}
Finally, using the real part of eqn (\ref{symmetry})
\begin{equation}
    \small
    \begin{aligned}
      \Re \{ \mathbf{x}^H \mathbf{K}_r \mathbf{x} \} = \mathbf{x}_1^T \Re&(\mathbf{K}_r)\mathbf{x}_1 ~+~ \mathbf{x}_2^T \Re(\mathbf{K}_r)\mathbf{x}_2 \\ ~+~ &j\mathbf{x}_1^T \Re(\mathbf{K}_r)\mathbf{x}_2 ~-~ j\mathbf{x}_2^T \Re(\mathbf{K}_r)\mathbf{x}_1 \\
      =~ &\mathbf{x}^{H}\Re\{ \mathbf{K}_{r}\}~\mathbf{x} .
    \end{aligned}
\end{equation}
This concludes the proof. \qedsymbol
\end{appendices}
\vspace{-0.3cm}
\medskip
\renewcommand*{\bibfont}{\footnotesize}
\printbibliography
\end{document}